\documentclass[%
 aip,
 pof,%
 amsmath,amssymb,
 reprint,%
]{revtex4-1}

\usepackage{graphicx}
\usepackage{dcolumn}
\usepackage{bm}
\usepackage[mathlines]{lineno}
\usepackage{color}
\usepackage{hyperref}
\begin{document}

\preprint{AIP/123-QED}

\title{Flow structure of compound droplets moving in microchannels}

\author{Zhizhao Che}
\email{chezhizhao@tju.edu.cn}
 \affiliation{State Key Laboratory of Engines, Tianjin University, Tianjin, 300072, China.}
\author{Yit Fatt Yap}%
\affiliation{%
Department of Mechanical Engineering, The Petroleum Institute, Abu Dhabi, UAE.
}%
\author{Tianyou Wang}
\affiliation{State Key Laboratory of Engines, Tianjin University, Tianjin, 300072, China.}%

\date{\today}

\begin{abstract}
Compound droplets can be used in substance encapsulation and material compartmentalization to achieve a precise control over the relevant processes in many applications, such as bioanalysis, pharmaceutical manufacturing, and material synthesis. The flow fields in compound droplets directly affect the performance of these applications, but it is challenging to measure them experimentally. In this study, the flow in compound droplets in axisymmetric microchannels is simulated using the Finite Volume Method, and the interface is captured using the Level Set Method with surface tension accounted for via the Ghost Fluid Method. The combination of the Level Set Method and the Ghost Fluid Method reduces spurious currents that are produced unphysically near the interface, and achieves a precise simulation of the complex flow field within compound droplets. The shape of compound droplets, the vortical patterns, the velocity fields, and the eccentricity are investigated and the effects of the key dimensionless parameters, including the size of the compound droplet, the size of the core droplet, the capillary number, and the viscosity ratio, are analyzed. The flow structures in multi-layered compound droplets are also studied. This study not only unveils the complex flow structure within compound droplets moving in microchannels, but can also be used to achieve a precise control over the relevant processes in a wide range of applications of compound droplets.
\end{abstract}


\maketitle


\section{Introduction}\label{sec:sec1}
Compound droplets are droplets with smaller droplets inside themselves forming core-shell structures. The shell of compound droplets can serve as a protective layer of the inner phase by minimizing the mass transfer between the core and the outer phase. Therefore, many functions can be achieved by loading the core with different substances, such as chemical reactants, bacteria, cells, pesticides, drugs, nutrients, and antibodies \cite{Wang2014DoubleEmulsion}. The compartmentalization of materials within core droplets allows a better control over the relevant processes. Hence, compound droplets play an important role in many applications, such as bioanalysis, pharmaceutical manufacturing, and material synthesis \cite{Datta2014DoubleEmulsionReview, Teh2008}.

The traditional method to produce compound droplets is to form a large amount of compound droplets in bulk, such as by shear generated by mechanical agitation. It results in compound droplets with wide size distributions. In contrast, with the development of micro-fabrication techniques during the past two decades, microfluidics offers an alternate route to produce monodisperse compound droplets one by one, and the droplet properties can be tuned precisely \cite{Datta2014DoubleEmulsionReview}. In microfluidics, compound droplets can be produced by forming the inner and the outer droplets subsequently at two droplet formation units \cite{Nisisako2005, Okushima2004, Deng2011LCWet, Wan2008, Che2012CPD, Che2017compound} (such as two T-junctions \cite{Nisisako2005, Okushima2004} or two flow-focusing geometries \cite{Deng2011LCWet, Wan2008}) or by forming the inner and the outer droplets simultaneously by properly combining two droplet formation units (such as a microcapillary structure proposed by Utada et al.\ \cite{Lorenceau2005, Utada2005}). To understand the formation of compound droplets, several numerical studies have been carried out to simulate the formation process in different microchannel structures \cite{Azarmanesh2016DoubleEmulsionGerris, Zhou2006, nabavi2015doubleCFD}, and the effects of relevant parameters have been reported, such as the geometry of the microfluidic device and the flow rates of different phases.

Even though many studies have been dedicated to the formation of compound droplets, the flow fields in compound droplets in microfluidics, to the best of our knowledge, have not be studied in details. Knowing the flow fields in compound droplets is of great significance because the flow fields directly affect the mass transfer in compound droplets and affect the insulation performance of the shell layer in the aforementioned applications, further leading to a better understanding of the mechanism and guiding the applications. Direct measurement of the flow fields in compound droplets in microfluidics is very difficult, not only because intrusive methods will disturb the complex flow structure in micrometer scales in the microfluidic devices, but also because the two layers of the interfaces pose challenges in optical measurement techniques including Particle Image Velocimetry (PIV), Particle Tracking Velocimetry (PTV), and Laser Induced Fluorescence (LIF) techniques due to optical distortion at the interfaces. In contrast to the obstacles encountered in experimental investigations, Computational Fluid Dynamics (CFD) shows its potential in analyzing the details of the complex flows in compound droplets.
Even though the flows of compound droplets in unconfined flows have been simulated, such as in linear flow \cite{Stone1990CompoundLinearFlow}, in extensional flow \cite{Qu2012CompoundExtensionalFlow}, and in shear flow \cite{Hua2014CompoundShearFlow, Smith2004CompoundShearFlowPRL}, they cannot describe the flows in microfluidics, since they are significantly different from those in microchannels/microcapillaries due to the confinement of the wall.
Zhou et al.\ \cite{ Zhou2008CompoundContraction} and Tao et al.\ \cite{Tao2013CompoundContraction} simulated the flow of compound droplets through contraction geometries. Song et al.\ \cite{Song2010PoFCompound} assumed the compound droplet is spherical and theoretically analyzed the flow of compound droplets in microchannels in low Reynolds number Stokes flow regime. In many microfluidics applications, the confinement by the wall is significant and the deformation of the compound droplet is not negligible.
To understand the flow details of compound droplets in microchannels, and unveil the confining effect of the wall on the deformation of the compound droplets and on the vortex patterns in compound droplets, systematic numerical studies on the flow fields in compound droplets in microfluidics are necessary for the development of the relevant applications.

This study aims to unveil the complex flow structures within compound droplets moving in microchannels. Given the small size of compound droplets, surface tension force dominates the flow. Therefore, accurate modeling of surface tension effect is the key to successful simulation of the complex flow fields in compound droplets. To achieve this, the Level Set Method (LSM) is used for interface capturing coupled with the Ghost Fluid Method (GFM) for surface tension implementation. The LSM offers an accurate calculation of the interface curvature. GFM incorporates the surface tension force in the momentum equation with minimized spurious current. The remainder of this paper is organized as follows. The numerical procedure is introduced in Section \ref{sec:sec2}, including the Finite Volume Method (FVM) for flow field, the LSM for interface capturing, and the GFM for surface tension. The results are presented and discussed in Section \ref{sec:sec3}, including the shape of compound droplets, the vortical structure, the eccentricity, the saddle points, etc. The effects of key controlling parameters are analyzed, and the flow fields within multiple layered compound droplets are also studied. Finally, conclusions are drawn in Section \ref{sec:sec4}.

\section{Numerical method}\label{sec:sec2}
\subsection{System configuration}\label{sec:sec21}
\begin{figure}[tb]
\centering
  \includegraphics[width=0.8\columnwidth]{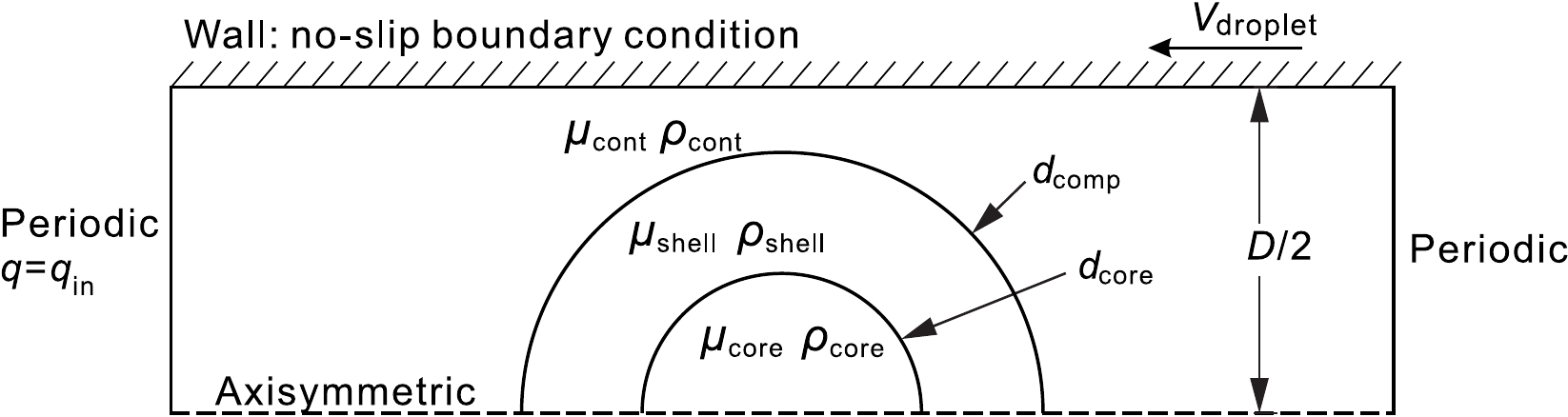}
  \caption{Schematic configuration of the numerical simulation of a compound droplet in an axisymmetric microchannel. }
  \label{fig:fig01}
\end{figure}

The simulation domain is a section of an axisymmetric microchannel with an inner diameter of $D$, as shown in FIG.\ \ref{fig:fig01}. The diameter of the compound droplet is ${{d}_{\text{comp}}}$ and the diameter of the core droplet is ${{d}_{\text{core}}}$. The viscosities and the densities are ${{\mu }_{\text{cont}}}$ and ${{\rho }_{\text{cont}}}$ for the continuous phase, ${{\mu }_{\text{shell}}}$ and ${{\rho }_{\text{shell}}}$ for the shell phase, and ${{\mu }_{\text{core}}}$ and ${{\rho }_{\text{core}}}$ for the core phase, respectively. The properties of the core fluid are set to be identical to the continuous phase in this study because this is the case for most applications because the shell layer is often used to isolate the core fluid and the continuous phase, such as water-in-oil-in-water (W/O/W) and oil-in-water-in-oil (O/W/O) compound droplets \cite{Okushima2004, Lorenceau2005, Utada2005, Nisisako2005, Deng2011LCWet, Che2017compound}. For high-level compound droplets, i.e., compound droplets with more than two layers, the fluids in different layers appear alternately \cite{Chu2007HighEmulsion, Wang2011HighEmulsion,Abate2009HighEmulsion,Kim2011HighEmulsion}.

In the simulation, no-slip boundary condition is specified on the wall of the microchannel, and the continuous phase fully wets the wall. This is consistent with most real applications of droplets in microchannels. In real droplet applications, this is achieved on purpose by selecting proper continuous fluids or surface modification because, if the droplet touches the wall, the droplet may stick on the wall and negatively affect the stability of the flow and the system performance \cite{Teh2008,Ralf2012}. The microchannel in the simulation is set to be long enough that it does not affect the flow in the compound droplets. A periodic boundary condition is used with a frame of reference following the droplet. In the periodic boundary condition, the velocity components are continuous at the left and the right boundaries while the average velocity is fixed to a constant, and the pressure field is handled through the SIMPLER algorithm \cite{Patankar1980}. Therefore, the pressure field condition is satisfied automatically since the velocity boundary condition has been imposed periodically. In addition, since the microchannel in the simulation is long and the droplet is in the center of the domain, the error introduced by the periodic boundary condition can be minimized.

The droplet speed is used to update the translating speed of the frame of reference, and it is calculated based on the mass of the shell phase, $v_\text{droplet}=\int\limits_{\Omega }{{v_\text{abs}}Hd\Omega }$, where $v_\text{abs}$ is the absolute local velocity of the fluid and $H$ is the smoothed Heaviside function defined in Eq.\ (\ref{eq:eq09}). Initially, the compound droplets are set to be concentric. They deform when flowing in the microchannel and gradually reach a steady state in the translating frame of reference. This study focuses on the flow after the compound droplets reach the steady state.

The controlling parameters for the flow of compound droplets in microchannels can be summarized into several dimensionless groups, which are used in simulations and analysis. The viscosity ratio is defined as the ratio of the viscosities between the shell phase and the core phase, $\hat{\mu } \equiv {{{\mu }_{\text{shell}}}}/{{{\mu }_{\text{core}}}}$. The density ratio is defined as the ratio of the densities between the shell phase and the core phase, $\hat{\rho } \equiv {{{\rho }_{\text{shell}}}}/{{{\rho }_{\text{core}}}}$. The size of the compound droplets is normalized by the diameter of the microchannel, ${{\hat{d}}_{\text{comp}}} \equiv {{{d}_{\text{comp}}}}/{D}$ , and the size of the core is normalized by the size of the compound droplet, ${{\hat{d}}_{\text{core}}} \equiv {{{d}_{\text{core}}}}/{{{d}_{\text{comp}}}}$. The capillary number is defined as the ratio between the viscous force and the surface tension force, $\text{Ca} \equiv {{{\mu }_{\text{cont}}}V}/{\sigma }$, where $V$ is the mean flow velocity in the microchannel, and ${{\mu }_{\text{cont}}}$ is the viscosity of the continuous phase. The Weber number is defined as the ratio between the inertia and the surface tension force, $\text{We} \equiv {{{\rho }_{\text{cont}}}{{V}^{2}}D}/{\sigma }$. These dimensionless numbers are not equally important. In most microfluidic devices, the inertia is usually very small due to the small size of the devices and the low flow speeds, i.e., $\text{We}\ll1$. In addition, for many liquids under nearly standard conditions of temperature and pressure, the densities are similar, i.e., $\hat{\rho } \sim 1$. Therefore, the effects of all dimensionless groups, except We and $\hat{\rho }$, are analyzed in this study.

\subsection{Finite Volume Method for flow fields}\label{sec:sec22}

All fluids considered in this study are incompressible and Newtonian. The FVM \cite{Che2011Pendent, Patankar1980} was used to discretize the continuity equation and the momentum equation in a cylindrical coordinate system,
\begin{equation}\label{eq:eq01}
  \frac{\partial \rho }{\partial t}+\frac{1}{r}\frac{\partial }{\partial r}\left( r\rho {{u}_{r}} \right)+\frac{\partial }{\partial z}\left( \rho {{u}_{z}} \right)=0,
\end{equation}
\begin{equation}\label{eq:eq02}
  \frac{\partial \left( \rho {{u}_{r}} \right)}{\partial t}+\frac{1}{r}\frac{\partial \left( r\rho {{u}_{r}}{{u}_{r}} \right)}{\partial r}+\frac{\partial \left( \rho {{u}_{z}}{{u}_{r}} \right)}{\partial z}
  =-\frac{\partial p}{\partial r}+\frac{1}{r}\frac{\partial }{\partial r}\left( \mu r\frac{\partial {{u}_{r}}}{\partial r} \right)+\frac{\partial }{\partial z}\left( \mu \frac{\partial {{u}_{r}}}{\partial z} \right)-\mu \frac{{{u}_{r}}}{{{r}^{2}}},
\end{equation}
\begin{equation}\label{eq:eq03}
  \frac{\partial \left( \rho {{u}_{z}} \right)}{\partial t}+\frac{\partial \left( \rho {{u}_{r}}{{u}_{z}} \right)}{\partial r}+\frac{\partial \left( \rho {{u}_{z}}{{u}_{z}} \right)}{\partial z}
  =-\frac{\partial p}{\partial z}+\frac{1}{r}\frac{\partial }{\partial r}\left( \mu r\frac{\partial {{u}_{z}}}{\partial r} \right)+\frac{\partial }{\partial z}\left( \mu \frac{\partial {{u}_{z}}}{\partial z} \right).
\end{equation}
The fluid properties $\rho$ and $\mu$ in each control volume is calculated using a smoothed Heaviside function $H$ as follows,
\begin{equation}\label{eq:eq04}
  \rho =H{{\rho }_{\text{shell}}}+\left( 1-H \right){{\rho }_{\text{core}}},
\end{equation}
\begin{equation}\label{eq:eq05}
  \frac{1}{\mu }=\frac{H}{{{\mu }_{\text{shell}}}}+\frac{1-H}{{{\mu }_{\text{core}}}}.
\end{equation}
The exact form of $H$ will be presented in Section \ref{sec:sec23}. The GFM \cite{Kang2000, Liu2000} accounts for  the surface tension through the pressure term,
\begin{equation}\label{eq:eq06}
  [p]=-\kappa \sigma,
\end{equation}
where $[p]$ indicates the Laplace pressure jump across the interface and $\kappa \equiv \nabla \cdot \mathbf{n}$ is the curvature of the interface. The symbol $\mathbf{n}$ denotes the unit normal vector to the interface, which is calculated in Eq.\ (\ref{eq:eq10}).

\subsection{Level Set Method for interface capturing}\label{sec:sec23}
The evolution of the droplet interface is captured using the LSM \cite{Osher2003}. The level set function $\phi$ is a signed distance from the interface. It is positive in one phase and negative in the other phase. The contour of $\phi=0$ represents the interface. The level set function $\phi$ is advected by the flow field obtained from the momentum equation,
\begin{equation}\label{eq:eq07}
  \frac{\partial \phi }{\partial t}+\mathbf{u}\cdot \nabla \phi =0.
\end{equation}
After evolving the level set function for several time steps, it generally ceases to remain as a signed distance function. Therefore, it is re-initialized through the re-initialization equation as
\begin{equation}\label{eq:eq08}
  \frac{\partial \phi }{\partial \tau }=\text{sign}(\phi)(1-|\nabla \phi|),
\end{equation}
where $\tau$ is the pseudo-time for the re-initialization. The spatial terms in Eqs.\ (\ref{eq:eq07}) and (\ref{eq:eq08}) are discretized using the fifth-order Weighted Essentially Non-Oscillatory (WENO) scheme \cite{Osher2003}, and the temporal terms are integrated using the third-order Runge-Kutta (RK) scheme with the Total Variation Diminishing (TVD) property \cite{Shu1988RK3}.
The smoothed Heaviside function $H$ in Eqs.\ (\ref{eq:eq04}) and (\ref{eq:eq05}) is defined as follows,
\begin{equation}\label{eq:eq09}
  H\equiv \left\{ \begin{matrix}   0, & {} & \phi <-\epsilon_\Delta   \\
   \frac{\phi +\epsilon_\Delta }{2\epsilon_\Delta }+\frac{1}{2\pi }\sin \left( \frac{\pi \phi }{\epsilon_\Delta } \right), & {} & -\epsilon_\Delta \leq \phi \leq \epsilon_\Delta   \\
   1, & {} & \phi >\epsilon_\Delta   \\
\end{matrix} \right.
\end{equation}
where $\epsilon_\Delta$ is set to be 1.5 times of the grid size.
The smoothed Heaviside function can smear the sharp interface into a belt region, approximate the abrupt jumps of fluid properties across the interface by gradual variations, and therefore, stabilize the simulation \cite{Osher2003, Tryggvason2011book}.
The curvature in Eq.\ (\ref{eq:eq06}) can be obtained from the level set function $\phi$ as follows,
\begin{equation}\label{eq:eq10}
\kappa \equiv \nabla \cdot \mathbf{n}=\nabla \cdot \frac{\nabla \phi }{\left| \nabla \phi  \right|}.
\end{equation}

It can be seen that the level set function, $\phi $, is a smooth function, being different from the volume fraction function in the volume of fluid method, which is a discontinuous function \cite{Tryggvason2011book}. Therefore, in the level set method, high-order discretization schemes can be easily implemented, such as ENO or WENO \cite{Osher2003}, and the curvature of the interface can also be easily calculated with high accuracy. One concern regarding the level method is the mass loss. To minimize the mass loss, besides the fifth-order WENO scheme and the third-order RK scheme, we added a local constraint term into Eq.\ (\ref{eq:eq08}) \cite{Osher2003}
\begin{equation}\label{eq1:eq10}
\frac{\partial \phi }{\partial \tau }=\text{sign}(\phi )(1-|\nabla \phi |)+\lambda \delta |\nabla \phi |,
\end{equation}
where $\delta $ is the smoothed delta function, which takes the form
\begin{equation}\label{eq1:eq11}
\delta \left( \phi  \right)\equiv \left\{ \begin{matrix}
   0, & {} & \phi <-\epsilon_\Delta   \\
   \frac{1}{2\epsilon_\Delta }+\frac{1}{2\epsilon_\Delta }\cos \left( \frac{\pi \phi }{\epsilon_\Delta } \right), & {} & -\epsilon_\Delta <\phi <\epsilon_\Delta   \\
   0, & {} & \phi >\epsilon_\Delta   \\
\end{matrix} \right.
\end{equation}
and
\begin{equation}\label{eq1:eq12}
{{\lambda }_{i,j}}=-\frac{\int_{{{\Omega }_{i,j}}}{\delta \left( \frac{{{\phi }^{n+1}}-{{\phi }^{n}}}{\Delta t} \right)d\mathbf{x}}}{\int_{{{\Omega }_{i,j}}}{{{\delta }^{2}}\left| \nabla \phi  \right|d\mathbf{x}}},
\end{equation}
where ${{\Omega }_{i,j}}$ refers to the control volume. By doing this, the mass loss relative to the initial mass in our simulation is always less than $10^{-4}$, from the initial condition to the final solution. It should also be noted that in level set simulations, mass loss tends to occur during the rapid evolution of the interface shape, which does not exist in our simulation.

\subsection{Validation of the simulations}\label{sec:sec24}
To check the accuracy and the validity of the present simulations, a series of rigorous tests are carried out, including the Zalesak's slotted disk rotation test for the accuracy of interface capturing, the spurious current, the mesh independence study, the effect of the initial location of the core droplet, and validation against experiment results. The details are provided in the Supplementary Materials. Among these tests, spurious currents are unphysical flows generated near the interface in the presence of surface tension. Their magnitudes should be much smaller than the characteristic velocity of the problem to avoid contaminating the actual flow field. Therefore, minimizing the spurious current is important in the simulations of the flow in compound droplets. In this study, the level set function is used to obtain the interface curvature with a high accuracy, and the GFM implements the surface tension force as a singular source term directly at the interface and avoid force imbalance. Therefore, with these simulation methods, the spurious current can be minimized in this study, as demonstrated in the Supplementary Materials.

\section{Results and discussion}\label{sec:sec3}
\subsection{Typical flow pattern in compound droplets}\label{sec:sec31}

\begin{figure}[tb]
\centering
  \includegraphics[width=\columnwidth]{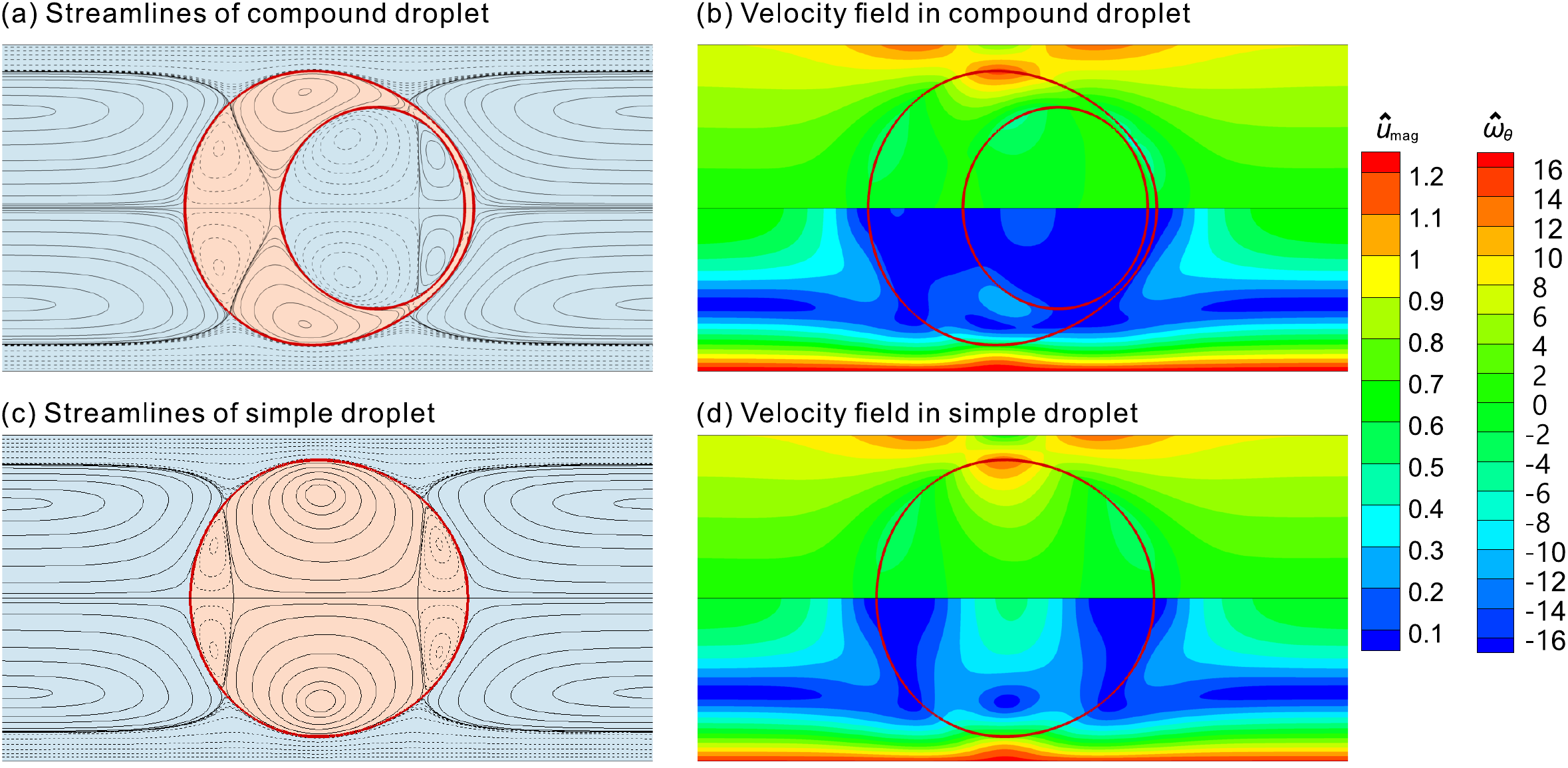}
  \caption{Flow field in a typical compound droplet in a microchannel and the comparison with that in a simple droplet. (a, c) Streamlines in the compound droplet and in the simple droplet, respectively; (b, d) Dimensionless vorticity (${{\hat{\omega }}_{\theta }}\equiv \frac{D}{V}\left( \frac{\partial {{u}_{r}}}{\partial z}-\frac{\partial {{u}_{z}}}{\partial r} \right)$, upper half of each figure) and dimensionless velocity magnitude (${\hat{u}_\text{mag}}\equiv {\sqrt{u_{r}^{2}+u_{z}^{2}}}/{V}$, lower half of each figure) in the compound droplet and in the simple droplet, respectively. The dimensionless parameters for the compound droplet and the simple droplet are ${{\hat{d}}_{\text{comp}}}={{\hat{d}}_{\text{simp}}}=0.85$, ${{\hat{d}}_{\text{core}}}=0.706$, $\text{Ca}=\text{0}\text{.0167}$, $\hat{\mu }=1$, $\hat{\rho }=1.19$, $\text{We}=\text{1}\text{.19}\times {{10}^{-7}}$. The solid streamlines have positive or zero stream function, indicating counter-clockwise recirculating direction, and the dashed streamlines have negative stream function, indicating clockwise recirculating direction. }
  \label{fig:fig02}
\end{figure}

The flow field in a typical compound droplet moving in a microchannel is shown in FIG.\ \ref{fig:fig02}.
In the steady state, the core droplet is in the frontal part of the compound droplet, resulting in a thin liquid shell in the front of the compound droplet and a thick shell in the rear. The core is only slightly deformed from a spherical shape by the flow. This is because the velocity in the core is very small, indicating a small capillary number and a relatively large effect of surface tension. The surface tension force tends to minimize the surface area of the core droplet forcing it to be more spherical. In contrast, the shell droplet is deformed more by the flow. The wall applies a strong shear force on the surface of the compound droplet, which pulls the rear of the compound droplet backward. Meanwhile, the front of the compound droplet is pushed forward by the spherical core, resulting in a spherical front of the compound droplet.

As the compound droplet moves in the microchannel, and due to the core-shell structure, complex vortical patterns form in the compound droplet. As the flow is axisymmetric, we discuss only the flow in the upper half of the figure. Since a moving frame of reference following the droplet is used, the droplet position is stationary in the moving frame of reference and the wall is moving backward, as shown in FIG.\ \ref{fig:fig02}a. In the core droplet, there are two vortices: one counter-clockwise rotating in the front and one clockwise rotating in the rear. The vortical pattern in the shell layer shares some similarity with that of a simple droplet, even though it is significantly affected by the core droplet. In a simple droplet under the same flow condition (see FIG.\ \ref{fig:fig02}c), there are three vortices, including a large vortex in the middle and two smaller vortices in the rear and the front. The large vortex is recirculating in the counter-clockwise direction, while the two small vortices are in the clockwise direction. In contrast, in the compound droplet, the presence of the core droplet effectively compresses the frontal vortex to within the thin shell layer, reduces the size of the middle vortex, and enlarges the rear vortex to a size comparable to the middle vortex.

Saddle points are important parameters to quantify the flow field. They are not only the points of flow separation, but also the point separating vortices and the location where the different streams meet. Therefore, it significantly affects the transportation of materials (such as reagent, surfactant, and particles) and the insulation performance of the shell in relevant applications. There are four saddle points on the outer surface of the compound droplet (two points with flow towards the interface and two points with flow from the interface), and three saddle points on the inner surface (two points with flow towards the interface and one point with flow from the interface).

The velocity magnitude within the compound droplet is shown in FIG.\ \ref{fig:fig02}b. The velocity magnitude within the droplet is small overall in the translating frame of reference following the droplet, much smaller than the translating speed of the wall. The velocity is also much smaller than that in a simple droplet, as shown in FIG.\ \ref{fig:fig02}d for comparison. The velocity magnitude is large close to the wall of the channel, and it applies a strong shear stress on the compound droplet and causes the compound droplet to deform. In the core of the compound droplet, the velocity magnitude is relatively large near the center, because the fluid near the center accelerates as it moves along the centerline. Similarly, a region of large velocity also appears near the center of the shell region behind the core droplet. In addition, the velocity is large near the surface of the core droplet, which is because the wall shear stress produces a recirculating vortex in the shell.

The compound droplet is set concentric initially in the simulation. As the compound droplet moves in the microchannel, the fluid in the frontal part of the shell drains and gradually reaches an equilibrium thickness. Even though the frontal part of the shell is thin, it does not collapse immediately by the surface tension effect which tends to minimize the surface area and causes the shell to break up. This is consistent with many experimental observations that compound droplets are relatively stable in many applications \cite{Datta2014DoubleEmulsionReview, Wang2014DoubleEmulsion}. This can be explained by examining the details of the flow field. There is a flow in the $-r$ direction in the continuous phase in front of the compound droplet. This flow can shear the fluid in the frontal shell layer towards the axis of the microchannel, and balance the drainage effect. Even though there is also a flow in the radial direction in the core droplet adjacent to the front shell layer, the velocity magnitude is much smaller in the core droplet than in the continuous phase. Consequently, the frontal shell layer drains until it reaches an equilibrium thickness and the droplet reaches an equilibrium position correspondingly. Of course, when the droplet deformation is very large, the shell is bound to collapse, but it is beyond the scope of the current study where we focus on the steady state of compound droplets.

\subsection{Effect of compound droplet size}\label{sec:sec32}
\begin{figure}[!tb]
\centering
  \includegraphics[width=\columnwidth]{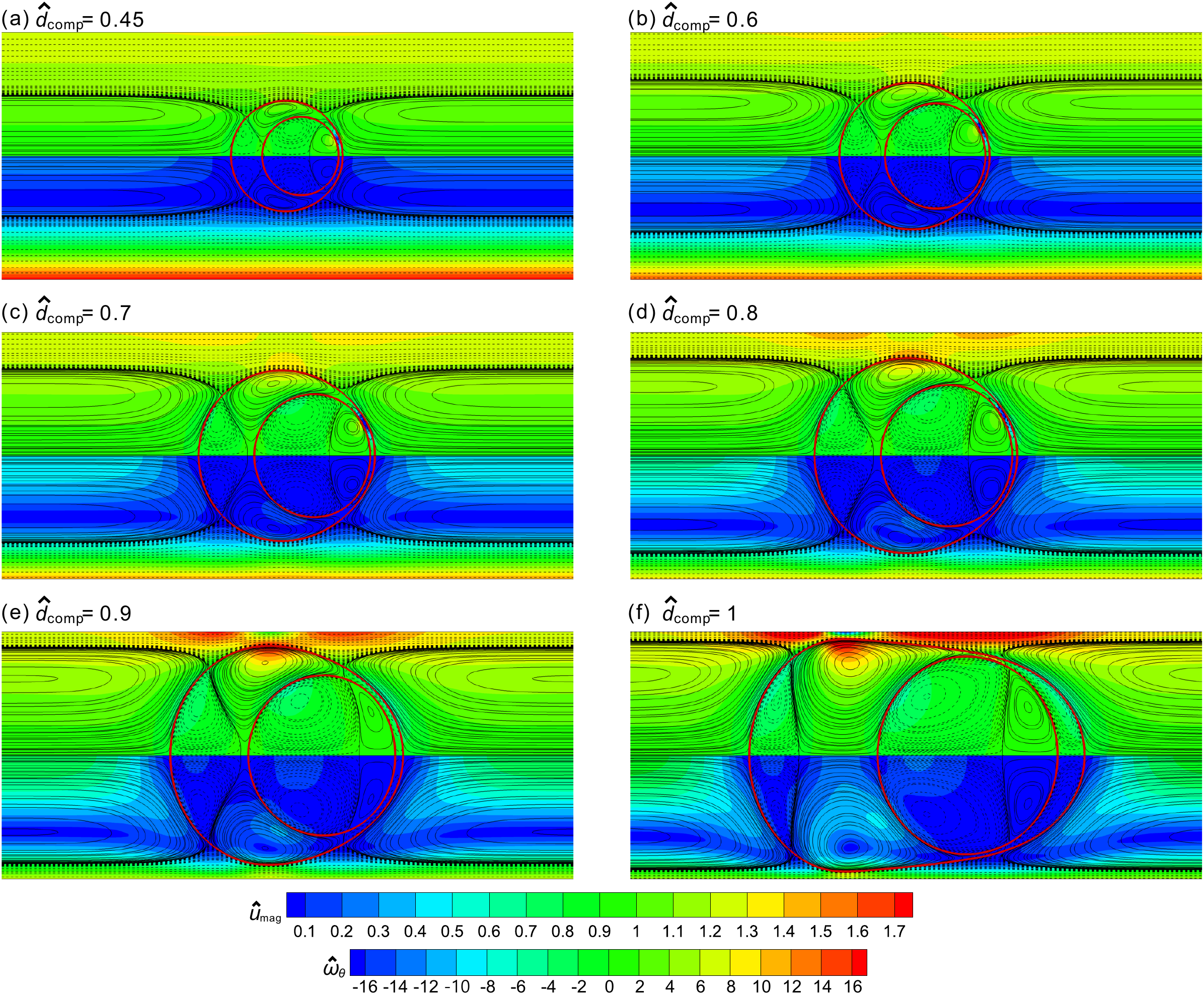}
  \caption{Effect of the size of compound droplets on the flow of compound droplets in microchannels. The other dimensionless parameters for the compound droplets are ${{\hat{d}}_{\text{core}}}=0.706$, $\text{Ca}=\text{0}\text{.0167}$, $\hat{\mu }=1$, $\hat{\rho }=1.19$, $\text{We}=\text{9}\text{.38}\times {{10}^{-8}}\sim \text{1}\text{.40}\times {{10}^{-7}}$. The color in the figures represents dimensionless vorticity (${{\hat{\omega }}_{\theta }}$, upper half of each figure) and dimensionless velocity magnitude ($\hat{u}_\text{mag}$, lower half of each figure), and the solid and dashed lines represent the streamlines. The solid streamlines have positive or zero stream function, indicating counter-clockwise recirculating direction, and the dashed streamlines have negative stream function, indicating clockwise recirculating direction.}
  \label{fig:fig03}
\end{figure}
The effect of the size of compound droplets are studied by varying the size of the compound droplets and fixing the other parameters including the relative size of the cores, as shown in FIG.\ \ref{fig:fig03}. For the compound droplets with a diameter much smaller than the diameter of the microchannel (see FIG.\ \ref{fig:fig03}a), only the flow near the center of the channel is affected by the presence of the compound droplet. The compound droplets move in the center of the channel at a much higher speed than the mean speed of the continuous fluid in the channel (represented by the high relative speed of the channel wall). The flow in the compound droplets is very weak, even though recirculating flow patterns are produced in the core and the shell of the compound droplets. Only a small portion of the fluid in the continuous phase is affected by the presence of the droplet, and most of the continuous fluid flows straightly in the direction along the microchannel.

With increasing the droplet size (see FIG.\ \ref{fig:fig03}c), more fluid in the microchannel is affected, and the speed of the droplet decreases. Visible velocity gradient appears both in the shell and in the core of the compound droplet. As a result, stronger recirculating flow forms in the compound droplet, owing to the increased shear stress exerted by the wall of the microchannel.

When the size of the compound droplet is comparable with the diameter of the microchannel (see FIG.\ \ref{fig:fig03}f), the flow in the compound droplet is significantly affected by the confining effect of the wall. The compound droplet occupies most of the cross section of the microchannel, and blocks most of the fluid in the continuous phase, resulting in a very thin layer of the continuous phase between the compound droplet and the wall. Consequently, the speed of the compound droplet is very close to the mean velocity of the fluid in the microchannel. The flow of the shell phase is severely retarded by the wall through the shear stress of the continuous phase, which results in large velocity gradients near the wall, indicating a large flow resistance of the compound droplet. As shown in FIG.\ \ref{fig:fig03}f, the compound droplet is significantly deformed by the flow of the continuous phase and the confining effect of the microchannel. The shell layer becomes very thick in the rear, and most shell fluid is in the rear of the compound droplet. The accumulation of the shell phase in the rear results in a large space for the flow development in the shell phase, and produces a relatively large velocity and consequently a strong recirculation in the shell.

\subsection{Effect of core size}\label{sec:sec33}

\begin{figure}[!tb]
\centering
  \includegraphics[width=0.8\columnwidth]{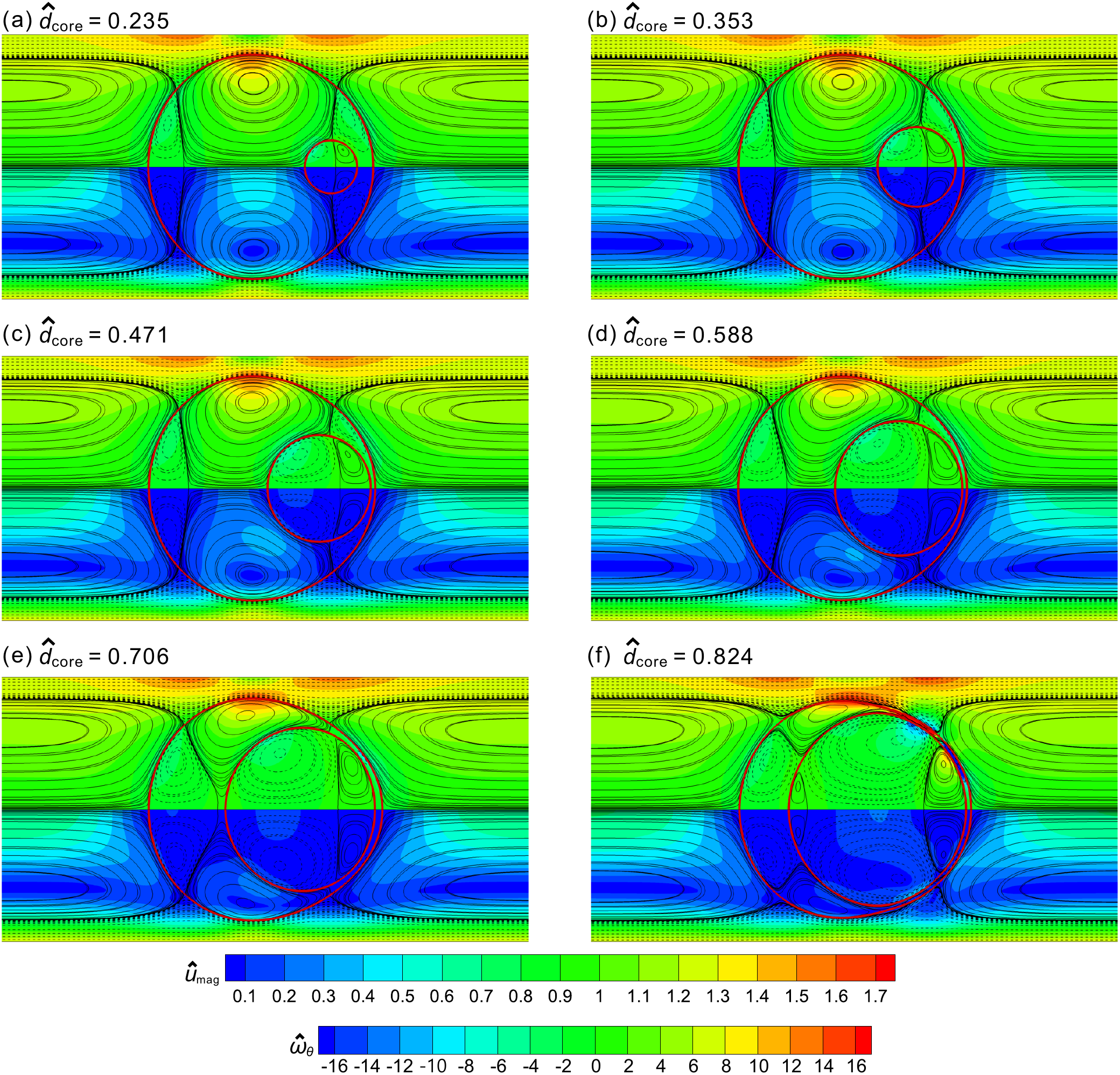}
  \caption{Effect of the core size on the flow patterns of compound droplets. The dimensionless parameters for the compound droplets are ${{\hat{d}}_{\text{comp}}}=0.85$, $\text{Ca}=\text{0}\text{.0167}$, $\hat{\mu }=1$, $\hat{\rho }=1.19$, $\text{We}=\text{1}\text{.19}\times {{10}^{-7}}$. The color in the figures represents dimensionless vorticity (${{\hat{\omega }}_{\theta }}$, upper half of each figure) and dimensionless velocity magnitude ($\hat{u}_\text{mag}$, lower half of each figure), and the solid and dashed lines represent the streamlines. The solid streamlines have positive or zero stream function, indicating counter-clockwise recirculating direction, and the dashed streamlines have negative stream function, indicating clockwise recirculating direction.}
  \label{fig:fig04}
\end{figure}

The effect of the core size of the compound droplet on the flow pattern is studied by varying the core size and fixing the other parameters (see FIG.\ \ref{fig:fig04}). With increasing the core size, the number of vortex in the core increases from two to three with the additional formation of a small vortex at the rear of the core. In addition, in the shell, the frontal vortex is significantly compressed by the core from a triangular shape to confined within a thin layer. As the core size increases further, the center vortex in the shell is also compressed into a wedged shape. However, the rear vortex in the shell increases its size due to the shrinkage of the middle vortex. With increasing the core size, the velocity magnitude in the core increases, and the velocity in the shell decreases. This is because the fluid in the core has more space to develop, and the fluid in the shell does not have enough space to develop.

\subsection{Effect of capillary number}\label{sec:sec34}

\begin{figure}[tb]
\centering
  \includegraphics[width=0.8\columnwidth]{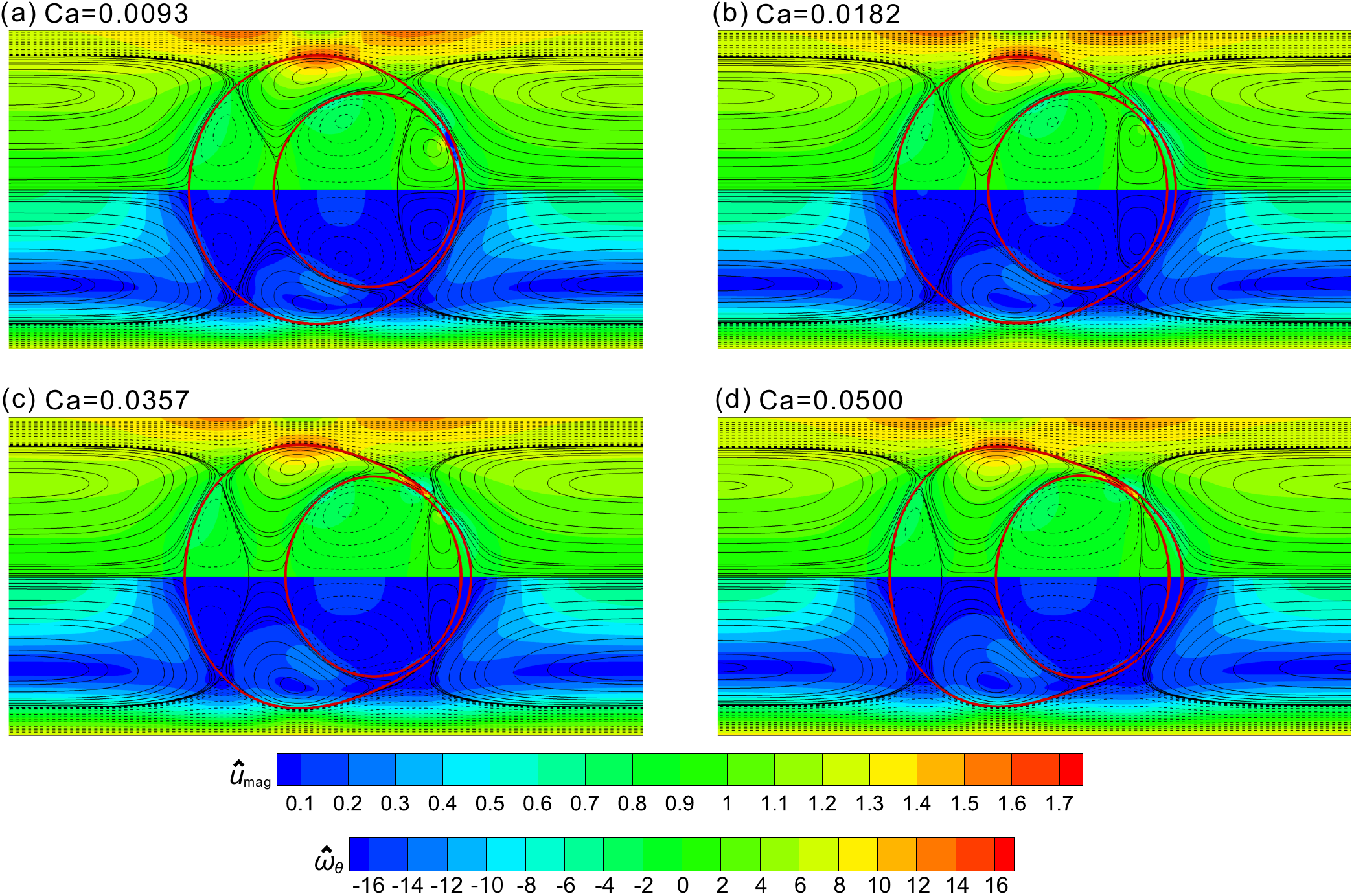}
  \caption{Effect of the capillary number on the flow of compound droplets in microchannels. The other dimensionless parameters for the compound droplets are ${{\hat{d}}_{\text{comp}}}=0.85$, ${{\hat{d}}_{\text{core}}}=0.706$, $\hat{\mu }=1$, $\hat{\rho }=1.19$, $\text{We}=\text{6}\text{.64}\times {{10}^{-8}}\sim \text{3}\text{.57}\times {{10}^{-7}}$. The color in the figures represents dimensionless vorticity (${{\hat{\omega }}_{\theta }}$, upper half of each figure) and dimensionless velocity magnitude ($\hat{u}_\text{mag}$, lower half of each figure), and the solid and dashed lines represent the streamlines. The solid streamlines have positive or zero stream function, indicating counter-clockwise recirculating direction, and the dashed streamlines have negative stream function, indicating clockwise recirculating direction.}
  \label{fig:fig05}
\end{figure}

The capillary number quantifies the relative effect between the viscous force and the surface tension force in the flow. Since many microfluidic devices of compound droplets operate in the regime of small capillary numbers to avoid significant deformation or even breakup of compound droplets, small Ca values ($\text{Ca}\leq0.05$) are considered here and they are varied by changing the surface tension of the fluid. As discussed in Section \ref{sec:sec21}, large surface tensions result in small capillary numbers. For a small capillary number, both the core and the shell droplets are almost spherical (see FIG.\ \ref{fig:fig05}a), indicating that the surface tension plays its role in restoring a spherical shape. As the capillary number increases, both the core and the shell droplets deform dramatically. The flow patterns in the droplets are also affected correspondingly. With increasing the capillary number (decreasing the surface tension), the frontal vortex in the core shrinks but the rear vortex expands. In addition, the main vortex in the shell fluid also expands, and the rear vortex shrinks. This is because the main vortex has more space to develop as the droplet deformation becomes larger at higher capillary numbers.

\subsection{Effect of viscosity ratio}\label{sec:sec35}
\begin{figure}[tb]
\centering
  \includegraphics[width=0.8\columnwidth]{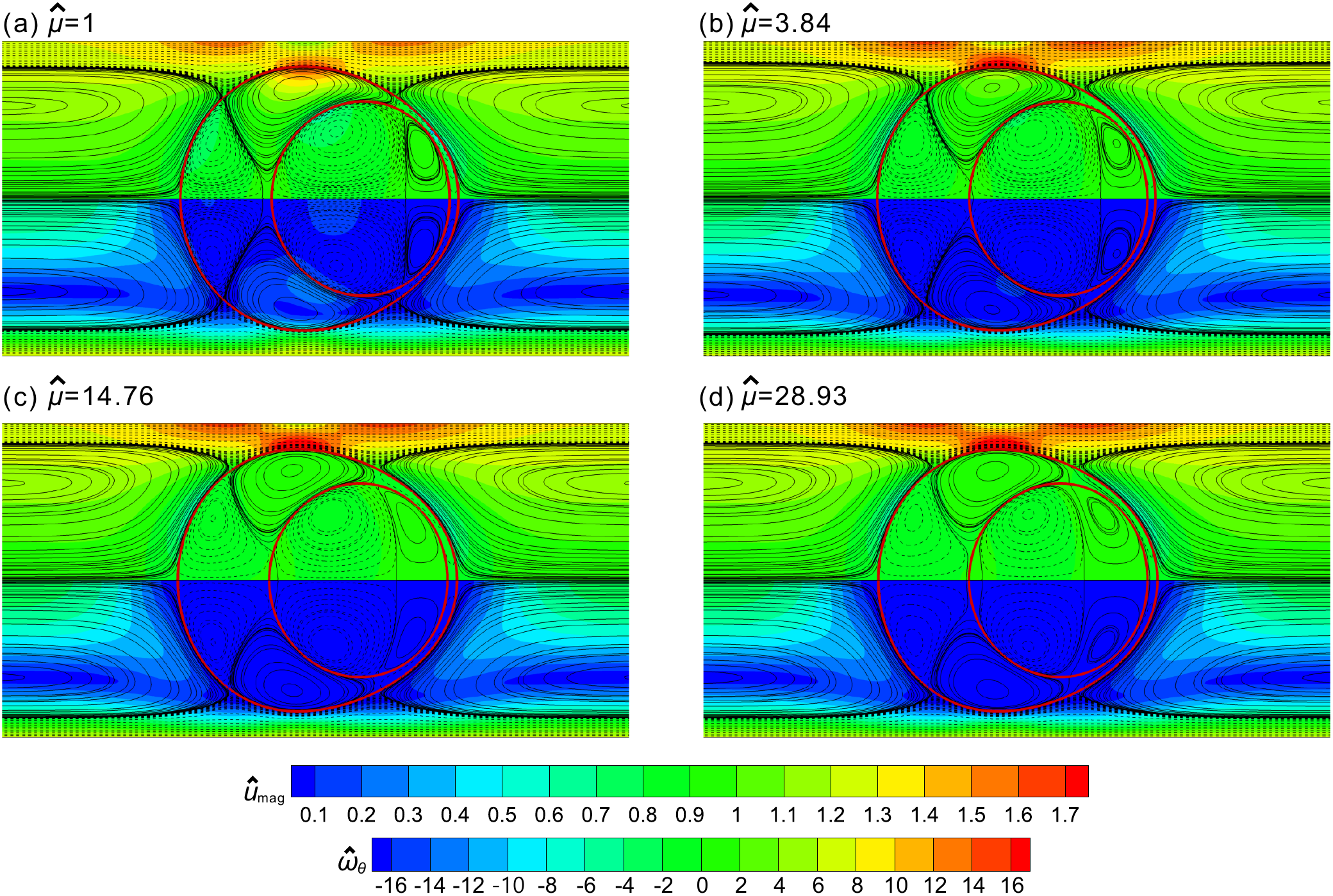}
  \caption{Effect of the viscosity ratio on the flow of compound droplets in microchannels. The dimensionless parameters for the compound droplets are ${{\hat{d}}_{\text{comp}}}=0.85$, ${{\hat{d}}_{\text{core}}}=0.706$, $\text{Ca}=\text{0}\text{.0167}$, $\hat{\rho }=1.19$, $\text{We}=\text{1}\text{.19}\times {{10}^{-7}}$. The color in the figures represents dimensionless vorticity (${{\hat{\omega }}_{\theta }}$, upper half of each figure) and dimensionless velocity magnitude ($\hat{u}_\text{mag}$, lower half of each figure), and the solid and dashed lines represent the streamlines. The solid streamlines have positive or zero stream function, indicating counter-clockwise recirculating direction, and the dashed streamlines have negative stream function, indicating clockwise recirculating direction.}
  \label{fig:fig06}
\end{figure}

The effect of the viscosity ratio is studied by changing the viscosity of the shell phase while keeping the viscosities of the continuous phase and the core phase unchanged, as shown in FIG.\ \ref{fig:fig06}. The viscosity of the continuous phase is not varied because changing it alters the capillary number of the flow. The shape of the compound droplet is not significantly affected upon increasing the viscosity ratio. This is because of the weak recirculating flow in the compound droplet. Thus, even though the shear stress in the shell is increased by increasing the shell viscosity, its effect is still much smaller than the surface tension force of the compound droplet. Therefore, the viscosity ratio does not affect the droplet shape significantly. However, it is worth noting that the flow pattern in the compound droplet is affected by the viscosity ratio. With increasing the viscosity of the shell, the rear vortex in the shell becomes larger, and compresses the middle vortex in the shell to shrink consequently.

\subsection{Eccentricity of compound droplets}\label{sec:sec36}
When compound droplets move in microchannels, the shell and the core are generally not concentric due to the internal recirculating flow in compound droplets. Eccentricity is an important feature of compound droplets, and can affect the application performance of compound droplets. For examples, compound droplets are often used as templates for the synthesis of particles with core/shell structures \cite{Dendukuri2009ParticleReview, Kumacheva2010PoymerizationReview}. A thin region of the shell of compound droplets will lead to a thin region in the shell of the particles, and influence the mechanical properties of the products significantly, such as the crush strength. In addition, considering that the shell of compound droplets can serve as a protective layer of the core fluid by minimizing the mass transfer between the core and the outer phases, a thin region of the shell will increase the undesirable transfer of species between the core and the continuous phases and deteriorate the insulation performance in relevant applications.

\begin{figure*}[!tb]
\centering
  \includegraphics[width=\columnwidth]{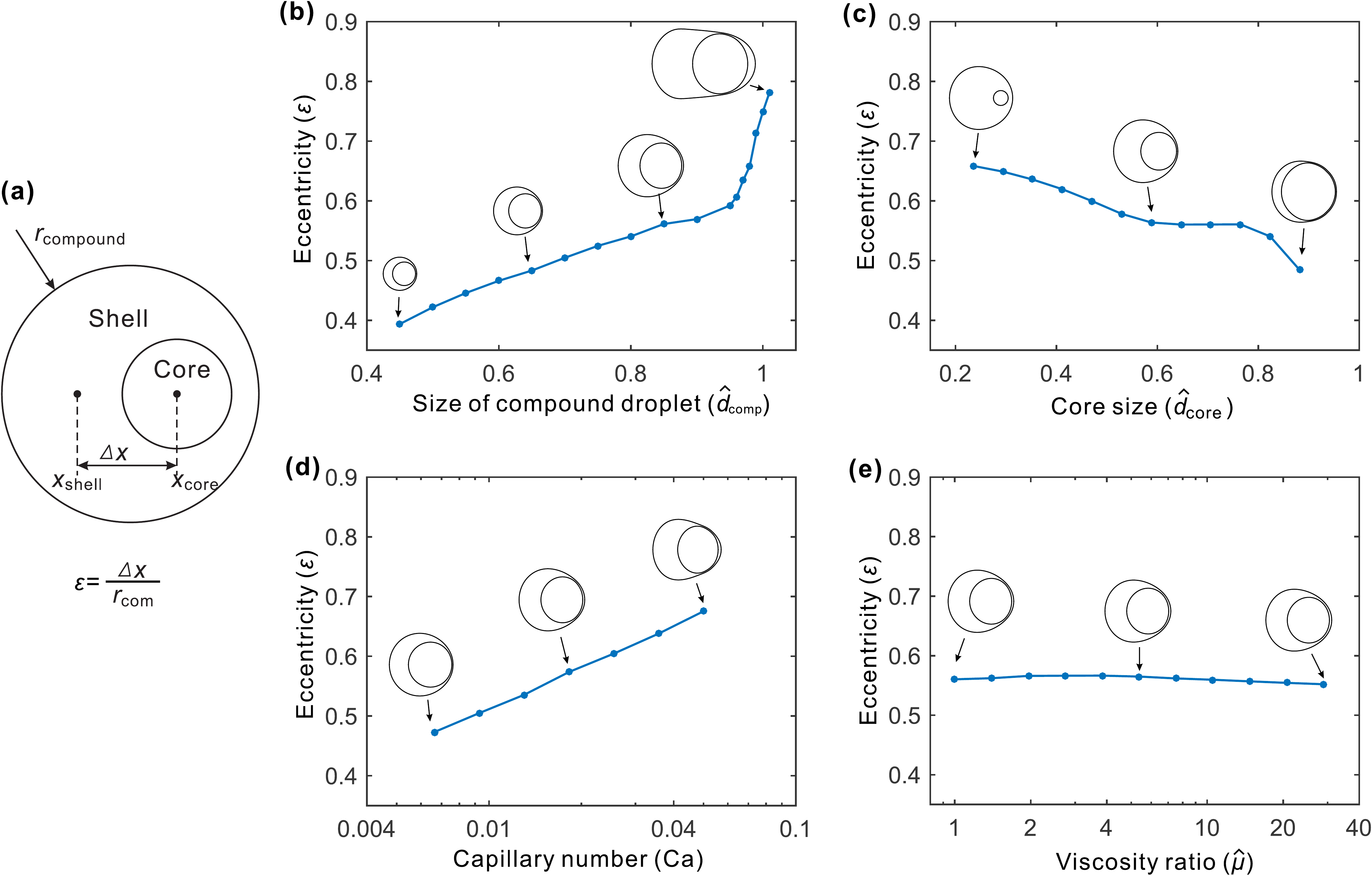}
  \caption{Eccentricity of compound droplets at different flow conditions in microchannels. (a) Eccentricity is defined to measure the shape of off-centered compound droplet: $\varepsilon ={\left| {{x}_{\text{core}}}-{{x}_{\text{shell}}} \right|}/{{{r}_{\text{comp}}}}$. (b) Effect of the size of compound droplets, ${{\hat{d}}_{\text{core}}}=0.706$, $\text{Ca}=\text{0}\text{.0167}$, $\hat{\mu }=1$, $\hat{\rho }=1.19$, $\text{We}=\text{9}\text{.38}\times {{10}^{-8}}\sim \text{1}\text{.40}\times {{10}^{-7}}$; (c) Effect of the core size, ${{\hat{d}}_{\text{comp}}}=0.85$, $\text{Ca}=\text{0}\text{.0167}$, $\hat{\mu }=1$, $\hat{\rho }=1.19$, $\text{We}=\text{1}\text{.19}\times {{10}^{-7}}$; (d) Effect of the capillary number, ${{\hat{d}}_{\text{comp}}}=0.85$, ${{\hat{d}}_{\text{core}}}=0.706$, $\hat{\mu }=1$, $\hat{\rho }=1.19$, $\text{We}=\text{6}\text{.64}\times {{10}^{-8}}\sim \text{3}\text{.57}\times {{10}^{-7}}$; (e) Effect of the viscosity ratio, ${{\hat{d}}_{\text{comp}}}=0.85$, ${{\hat{d}}_{\text{core}}}=0.706$, $\text{Ca}=\text{0}\text{.0167}$, $\hat{\rho }=1.19$, $\text{We}=\text{1}\text{.19}\times {{10}^{-7}}$.}
  \label{fig:fig07}
\end{figure*}

To quantify the shape of the compound droplet and the position of the core in the compound droplet, the eccentricity is defined as follows (see FIG.\ \ref{fig:fig07}a),
\begin{equation}\label{eq:eq11}
  \varepsilon \equiv \frac{\left| {{x}_{\text{core}}}-{{x}_{\text{shell}}} \right|}{{{r}_{\text{comp}}}},
\end{equation}
where ${{x}_{\text{core}}}$ and ${{x}_{\text{shell}}}$ are the centroids of the core droplet and the shell phase respectively, and ${{r}_{\text{comp}}}$ is the equivalent radius of the compound droplet calculated based on the volume of the compound droplet, ${{r}_{\text{comp}}} \equiv {{\left[ {3\left( {{V}_{\text{shell}}}+{{V}_{\text{core}}} \right)}/{\left( 4\pi  \right)} \right]}^{1/3}}$, where ${{V}_{\text{shell}}}$  and ${{V}_{\text{core}}}$ are the volumes of the shell and the core phases, respectively. The eccentricity of a concentric compound droplet is 0, and it increases as the core shifts towards the front of the compound droplet.

The eccentricity is a strong function of the compound droplet size, as shown in FIG.\ \ref{fig:fig07}b. With increasing the compound droplet size, eccentricity increases, and the effect becomes stronger for large compound droplets (${{\hat{d}}_{\text{comp}}}>0.9$). Small compound droplets are almost spherical and the deformation increases with the droplet size, as discussed in Section \ref{sec:sec32}. For compound droplets smaller than 0.9, the deformation is mainly due to the flow within and outside the compound droplets. When the compound droplet is larger than 0.9, the confining effect becomes dominant and the deformation increase dramatically with the droplet size. In contrast, the core is always less deformed than the shell. Therefore, with increasing the size of compound droplet, the eccentricity first increases slowly because of the increasing deformation by the flow, and then increases dramatically due to the confining effect of the wall.

With increasing the core size, the eccentricity of the compound droplet decreases, as shown in FIG.\ \ref{fig:fig07}c. This is because the core in equilibrium is in the front of the compound droplet. The equilibrium position of the core is mainly determined by the flow in the shell, in the core, and in front of the compound droplet. The resulted equilibrium shell thickness is always much smaller than the size of the compound droplet. Therefore, as the droplet size increases, the eccentricity of the compound droplet decreases.

Eccentricity is also affected by the capillary number of the flow (see FIG.\ \ref{fig:fig07}d). For small capillary numbers, the compound droplets are close to spherical due to the strong effect of surface tension, resulting in small eccentricity values. With increasing the capillary number, the effect of surface tension weakens, and viscous effect plays a more important role. Consequently the compound droplet shape is deformed significantly by the flow and most of the shell fluid is squeezed to the rear of the compound droplet, as explained in Section \ref{sec:sec34}. Therefore, the eccentricity of compound droplets increases with increasing the capillary number.

The eccentricity of compound droplet is not significantly affected by the viscosity ratio between the core and the shell phase, as shown in FIG.\ \ref{fig:fig07}e. This is because the flow in the compound droplet is relatively slow and dominated by the surface tension force. Even for large viscosity ratios, the relative effect of the viscosity is still much smaller than the surface tension effect. Thus, the eccentricity of the compound droplet does not change significantly with viscosity ratio.

\subsection{Aspect ratio of compound droplets}\label{sec:sec36a}
\begin{figure*}[!tb]
\centering
  \includegraphics[width=\columnwidth]{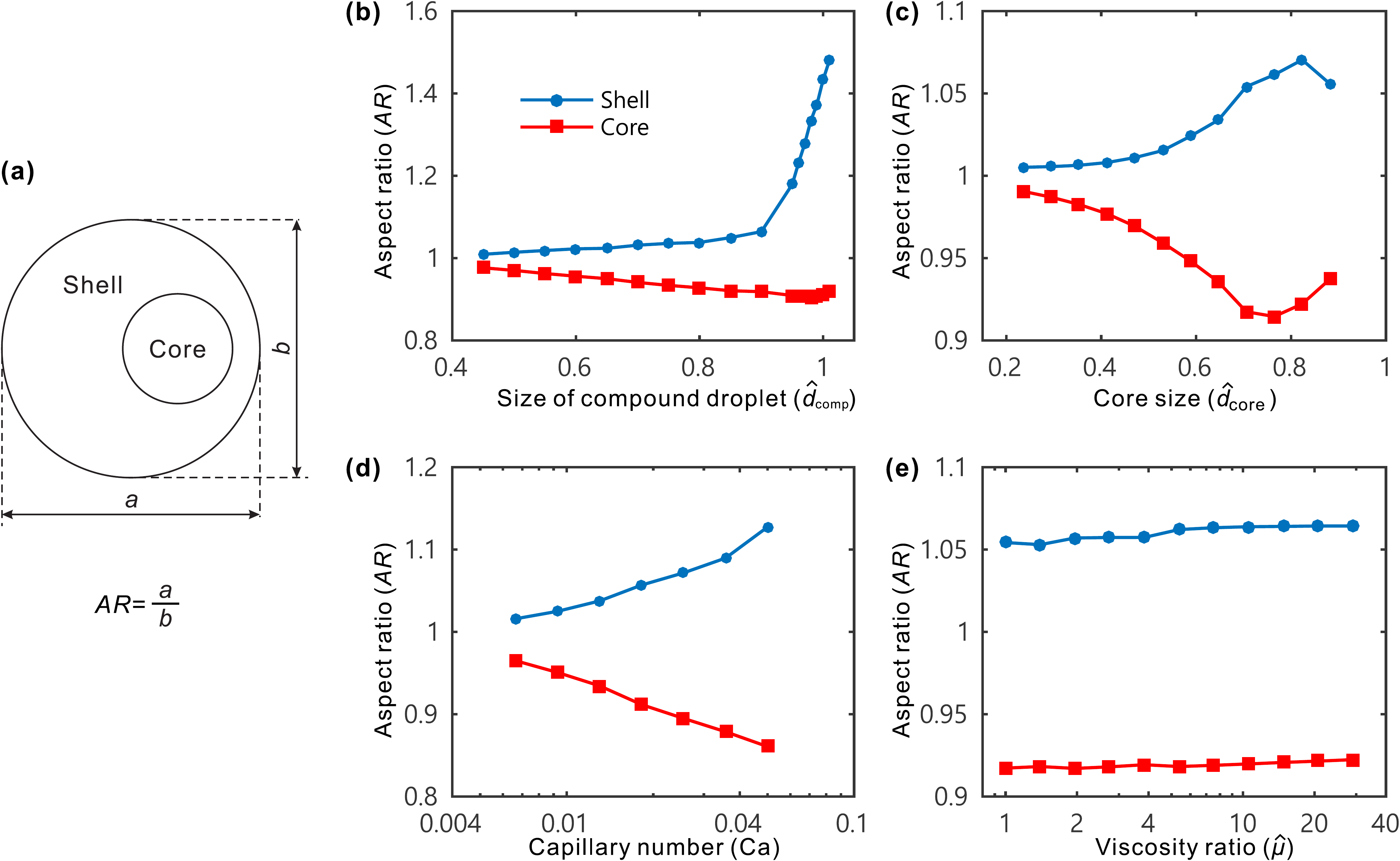}
  \caption{Aspect ratio of compound droplets. (a) The aspect ratio is defined as the ratio between the length of the droplet in the flow direction and that in the spanwise direction, i.e., $AR\equiv a/b$. (b) Effect of the size of compound droplets; (c) Effect of the core size; (d) Effect of the capillary number; (e) Effect of the viscosity ratio. The parameters are the same to those in FIG.\ \ref{fig:fig07}.}
  \label{fig:fig08}
\end{figure*}
The deformation of compound droplets is also quantified by the aspect ratio of the compound droplets, as defined in FIG.\ \ref{fig:fig08}a and the effects of the key controlling parameters are analyzed, as shown in FIG.\ \ref{fig:fig08}b-e.

For very small compound droplets, the aspect ratios of the shell and of the core are both close to unity, indicating that the droplet is spherical, as shown in FIG.\ \ref{fig:fig08}b. Since the flow in the shell pushes the core forwards behind the core, the aspect ratio of the core is smaller than unity. In contrast, since the core pushes the shell forwards in the front of the shell, the aspect ratio of the shell is larger than unity. Therefore, the shell is prolate while the core is oblate. As the size of the compound droplet increases, the aspect ratio of the shell increases slight, and that of the core decreases slightly. This is mainly due to the increased deformation of the compound droplet by the flow. As the dimensionless droplet size $\hat d_\text{comp}$ approaches unity, the aspect ratio of the shell increases dramatically, owing to the confinement effect of the wall. Correspondingly, the compound droplet becomes elongated, as shown in the snapshot in FIG.\ \ref{fig:fig07}b.

The effect of the core size of the compound droplet on the aspect ratio is shown in FIG.\ \ref{fig:fig08}c. With a small core, the core is almost spherical, exerting a small pushing force in the front of the shell. Therefore, the shell is also spherical  ($AR\approx1$). As the core size increases, the surface tension effect of the core decreases, the core becomes easy to deform, and $AR$ of the core decrease. Correspondingly, the core exerts a larger force to the shell, resulting a larger deformation of the shell and a larger $AR$.  As the core size approaches the shell size, the flow in the shell is inhibited by the small space in the shell, and surface tension forces of the shell and the core restore the compound droplet to spherical. Therefore, the aspect ratio of the shell decreases and that of the core increases.

The effect of the capillary number on the aspect ratio is shown in FIG.\ \ref{fig:fig08}d. With increasing the capillary number, the effect of the surface tension reduces, and the compound droplet becomes easier to deform. Therefore, the aspect ratio of the core decreases, and that of the shell increases. In addition, the aspect ratio of the compound droplet does not change significantly with the viscosity ratio, as shown in FIG.\ \ref{fig:fig08}e, since the flow in the shell is weak and variation in the shell viscosity does not significantly affect the droplet deformation, as shown in the snapshot of the compound droplets in FIG.\ \ref{fig:fig07}e.

\subsection{Multi-layered compound droplets}\label{sec:sec37}
High-order compound droplets are compound droplets with more layers. They can be used to achieve high level of encapsulation and compartmentalization \cite{Lee2016TripleWeitz} for complex reactions and analysis in various applications, including drug delivery, material synthesis, and bio-analysis. The formation of monodisperse high-order compound droplets have been achieved in microfluidics in glass microcapillary devices \cite{Chu2007HighEmulsion, Kim2011HighEmulsion, Lee2016TripleWeitz, Wang2011HighEmulsion} and in soft-lithographically fabricated PDMS devices \cite{Abate2009HighEmulsion}.

\begin{figure}[tb]
\centering
  \includegraphics[width=0.5\columnwidth]{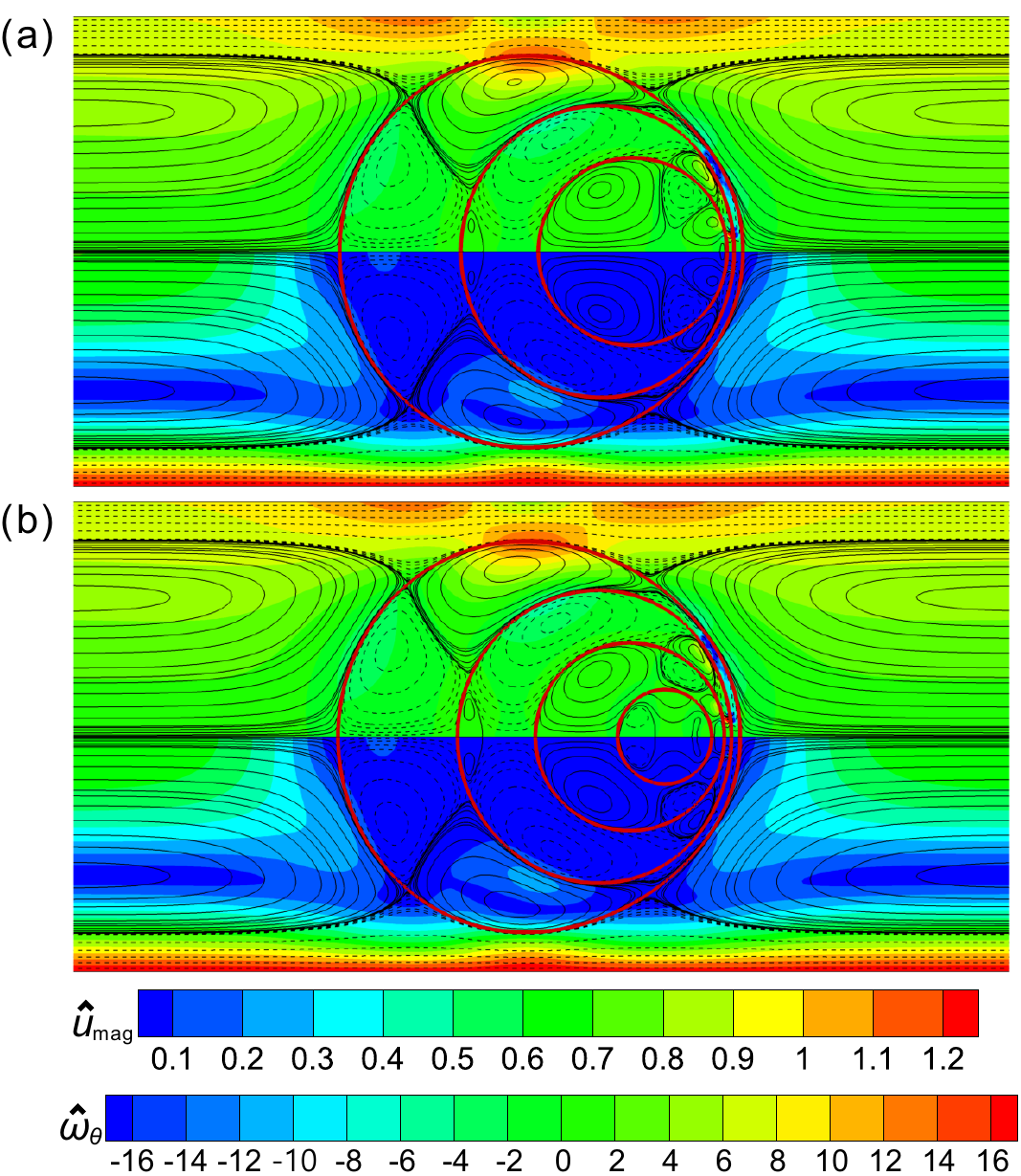}
  \caption{Flow patterns in multiple-layered compound droplets. (a) A triple compound droplet; (b) A quadruple compound droplet. The dimensionless parameters for the compound droplets are ${{\hat{d}}_{\text{comp}}}=0.8$, $\text{Ca}=\text{0}\text{.0167}$, $\hat{\mu }=1$, $\hat{\rho }=1.19$, $\text{We}=\text{1}\text{.19}\times {{10}^{-7}}$. The core sizes for the triple compound droplet are ${{\hat{d}}_{\text{core1}}}=0.6$ and ${{\hat{d}}_{\text{core2}}}=0.4$. The core sizes of the quadruple compound droplet are ${{\hat{d}}_{\text{core1}}}=0.6$, ${{\hat{d}}_{\text{core2}}}=0.4$, and ${{\hat{d}}_{\text{core3}}}=0.2$. The color in the figures represents dimensionless vorticity (${{\hat{\omega }}_{\theta }}$, upper half of each figure) and dimensionless velocity magnitude ($\hat{u}_\text{mag}$, lower half of each figure), and the solid and dashed lines represent the streamlines. The solid streamlines have positive or zero stream function, indicating counter-clockwise recirculating direction, and the dashed streamlines have negative stream function, indicating clockwise recirculating direction.}
  \label{fig:fig09}
\end{figure}

The flow fields in a triple compound droplet and a quadruple compound droplet are shown in FIG.\ \ref{fig:fig09}. Both are featured with complex recirculating patterns. The inner droplets are all in the frontal parts of the outer droplets, resulting in thicker shells in the frontal part than those in the rear part. These interfacial shapes severely compress the vortices in the frontal parts of the shells, and results in smaller vortices in the frontal parts than in the rear parts. In addition, these vortices are aligned in the radial direction of the compound droplets attributed to the interaction with the flow in the adjacent layers.

\section{Conclusions}\label{sec:sec4}
To understand the complex flow structure within compound droplets moving in axisymmetric microchannels, the flow is simulated using the Finite Volume Method and the interface is captured using the Level Set Method. The surface tension effect is implemented using the Ghost Fluid Method. With the combination of the Level Set Method and the Ghost Fluid Method, spurious current produced unphysically near the interface is minimized in the simulation, which is essential to simulate the complex flow structure of compound droplets.

In this study, the droplet shape, the vortical pattern, the flow velocity field, the eccentricity, and the effects of the key controlling parameters, such as the compound droplet size, the core size, the capillary number, and the viscosity ratio are analyzed. The results show that complex flow patterns form in the compound droplets due to the presence of the core droplets and the confining effect of the wall of the microchannel. Small compound droplets have weak internal recirculation and only affect the flow near the center of the microchannel, while large compound droplets experience severe deformation by the flow and the wall. With increasing the core size, the vortices in the shell is compressed and becomes weaker, and the vortices in the core expand and become stronger. With increasing the capillary number, the compound droplets are more deformed, and the eccentricity increases. The viscosity ratio does not significantly affect the droplet shape and the eccentricity. The flow in multi-layered compound droplets is analyzed, and the vortices in different layers are aligned in the radial direction due to the interaction with the flow in the adjacent layers.

The complex flow patterns in compound droplets can significantly affect not only the relevant transport phenomena, such as heat transfer \cite{Bandara2015DropletHeatTransferReview, Che2012HeatTransfer2DTA, Che2015HeatTransfer3Ddroplet} and mass transfer \cite{Handique2001, Tice2003DropletMixing, Che2011ChaoticMixingPRE}, but also the compartmentalization performance for substance encapsulation \cite{Wang2014DoubleEmulsion} in a wide range of applications. This study not only provides insight into the mechanism of the flow of compound droplets in microchannels/microcapillaries, but also can help guide the design of the applications and enable a precise control over the relevant processes. This study focuses only on the steady state of compound droplets. There are many open questions about compound droplets moving in microchannels. For example, compound droplets under large deformation may become unstable. Systematic investigation on their stability could provide more information on the dynamics of compound droplets.

\section*{Supplementary Material}
See supplementary material for the details of the validation of the numerical simulations, including the Zalesak's slotted disk rotation test for the accuracy of interface capturing (FIG.\ S1), the spurious current (FIG.\ S2), the mesh independence study (FIG.\ S3), the effect of the initial location of the core droplet (FIG.\ S4), and validation against experiment results (FIG.\ S5).

\section*{Acknowledgements}
This work is supported by National Natural Science Foundation of China (Grant No.\ 51676137) and Natural Science Foundation of Tianjin City (Grant No.\ 16JCYBJC41100).

\bibliography{CompoundDroplet}

\begin{thebibliography}{42}%
\makeatletter
\providecommand \@ifxundefined [1]{%
 \@ifx{#1\undefined}
}%
\providecommand \@ifnum [1]{%
 \ifnum #1\expandafter \@firstoftwo
 \else \expandafter \@secondoftwo
 \fi
}%
\providecommand \@ifx [1]{%
 \ifx #1\expandafter \@firstoftwo
 \else \expandafter \@secondoftwo
 \fi
}%
\providecommand \natexlab [1]{#1}%
\providecommand \enquote  [1]{``#1''}%
\providecommand \bibnamefont  [1]{#1}%
\providecommand \bibfnamefont [1]{#1}%
\providecommand \citenamefont [1]{#1}%
\providecommand \href@noop [0]{\@secondoftwo}%
\providecommand \href [0]{\begingroup \@sanitize@url \@href}%
\providecommand \@href[1]{\@@startlink{#1}\@@href}%
\providecommand \@@href[1]{\endgroup#1\@@endlink}%
\providecommand \@sanitize@url [0]{\catcode `\\12\catcode `\$12\catcode
  `\&12\catcode `\#12\catcode `\^12\catcode `\_12\catcode `\%12\relax}%
\providecommand \@@startlink[1]{}%
\providecommand \@@endlink[0]{}%
\providecommand \url  [0]{\begingroup\@sanitize@url \@url }%
\providecommand \@url [1]{\endgroup\@href {#1}{\urlprefix }}%
\providecommand \urlprefix  [0]{URL }%
\providecommand \Eprint [0]{\href }%
\providecommand \doibase [0]{http://dx.doi.org/}%
\providecommand \selectlanguage [0]{\@gobble}%
\providecommand \bibinfo  [0]{\@secondoftwo}%
\providecommand \bibfield  [0]{\@secondoftwo}%
\providecommand \translation [1]{[#1]}%
\providecommand \BibitemOpen [0]{}%
\providecommand \bibitemStop [0]{}%
\providecommand \bibitemNoStop [0]{.\EOS\space}%
\providecommand \EOS [0]{\spacefactor3000\relax}%
\providecommand \BibitemShut  [1]{\csname bibitem#1\endcsname}%
\let\auto@bib@innerbib\@empty
\bibitem [{\citenamefont {Wang}, \citenamefont {Zhang},\ and\ \citenamefont
  {Chu}(2014)}]{Wang2014DoubleEmulsion}%
  \BibitemOpen
  \bibfield  {author} {\bibinfo {author} {\bibfnamefont {W.}~\bibnamefont
  {Wang}}, \bibinfo {author} {\bibfnamefont {M.-J.}\ \bibnamefont {Zhang}}, \
  and\ \bibinfo {author} {\bibfnamefont {L.-Y.}\ \bibnamefont {Chu}},\
  }\bibfield  {title} {\enquote {\bibinfo {title} {Microfluidic approach for
  encapsulation via double emulsions},}\ }\href@noop {} {\bibfield  {journal}
  {\bibinfo  {journal} {Curr. Opin. Pharmacol.}\ }\textbf {\bibinfo {volume}
  {18}},\ \bibinfo {pages} {35--41} (\bibinfo {year} {2014})}\BibitemShut
  {NoStop}%
\bibitem [{\citenamefont {Datta}\ \emph {et~al.}(2014)\citenamefont {Datta},
  \citenamefont {Abbaspourrad}, \citenamefont {Amstad}, \citenamefont {Fan},
  \citenamefont {Kim}, \citenamefont {Romanowsky}, \citenamefont {Shum},
  \citenamefont {Sun}, \citenamefont {Utada},\ and\ \citenamefont
  {Windbergs}}]{Datta2014DoubleEmulsionReview}%
  \BibitemOpen
  \bibfield  {author} {\bibinfo {author} {\bibfnamefont {S.~S.}\ \bibnamefont
  {Datta}}, \bibinfo {author} {\bibfnamefont {A.}~\bibnamefont {Abbaspourrad}},
  \bibinfo {author} {\bibfnamefont {E.}~\bibnamefont {Amstad}}, \bibinfo
  {author} {\bibfnamefont {J.}~\bibnamefont {Fan}}, \bibinfo {author}
  {\bibfnamefont {S.-H.}\ \bibnamefont {Kim}}, \bibinfo {author} {\bibfnamefont
  {M.}~\bibnamefont {Romanowsky}}, \bibinfo {author} {\bibfnamefont {H.~C.}\
  \bibnamefont {Shum}}, \bibinfo {author} {\bibfnamefont {B.}~\bibnamefont
  {Sun}}, \bibinfo {author} {\bibfnamefont {A.~S.}\ \bibnamefont {Utada}}, \
  and\ \bibinfo {author} {\bibfnamefont {M.}~\bibnamefont {Windbergs}},\
  }\bibfield  {title} {\enquote {\bibinfo {title} {Double emulsion templated
  solid microcapsules: Mechanics and controlled release},}\ }\href@noop {}
  {\bibfield  {journal} {\bibinfo  {journal} {Adv. Mater.}\ }\textbf {\bibinfo
  {volume} {26}},\ \bibinfo {pages} {2205--2218} (\bibinfo {year}
  {2014})}\BibitemShut {NoStop}%
\bibitem [{\citenamefont {Teh}\ \emph {et~al.}(2008)\citenamefont {Teh},
  \citenamefont {Lin}, \citenamefont {Hung},\ and\ \citenamefont
  {Lee}}]{Teh2008}%
  \BibitemOpen
  \bibfield  {author} {\bibinfo {author} {\bibfnamefont {S.~Y.}\ \bibnamefont
  {Teh}}, \bibinfo {author} {\bibfnamefont {R.}~\bibnamefont {Lin}}, \bibinfo
  {author} {\bibfnamefont {L.~H.}\ \bibnamefont {Hung}}, \ and\ \bibinfo
  {author} {\bibfnamefont {A.~P.}\ \bibnamefont {Lee}},\ }\bibfield  {title}
  {\enquote {\bibinfo {title} {Droplet microfluidics},}\ }\href@noop {}
  {\bibfield  {journal} {\bibinfo  {journal} {Lab Chip}\ }\textbf {\bibinfo
  {volume} {8}},\ \bibinfo {pages} {198--220} (\bibinfo {year}
  {2008})}\BibitemShut {NoStop}%
\bibitem [{\citenamefont {Nisisako}, \citenamefont {Okushima},\ and\
  \citenamefont {Torii}(2005)}]{Nisisako2005}%
  \BibitemOpen
  \bibfield  {author} {\bibinfo {author} {\bibfnamefont {T.}~\bibnamefont
  {Nisisako}}, \bibinfo {author} {\bibfnamefont {S.}~\bibnamefont {Okushima}},
  \ and\ \bibinfo {author} {\bibfnamefont {T.}~\bibnamefont {Torii}},\
  }\bibfield  {title} {\enquote {\bibinfo {title} {Controlled formulation of
  monodisperse double emulsions in a multiple-phase microfluidic system},}\
  }\href@noop {} {\bibfield  {journal} {\bibinfo  {journal} {Soft Matter}\
  }\textbf {\bibinfo {volume} {1}},\ \bibinfo {pages} {23--27} (\bibinfo {year}
  {2005})}\BibitemShut {NoStop}%
\bibitem [{\citenamefont {Okushima}\ \emph {et~al.}(2004)\citenamefont
  {Okushima}, \citenamefont {Nisisako}, \citenamefont {Torii},\ and\
  \citenamefont {Higuchi}}]{Okushima2004}%
  \BibitemOpen
  \bibfield  {author} {\bibinfo {author} {\bibfnamefont {S.}~\bibnamefont
  {Okushima}}, \bibinfo {author} {\bibfnamefont {T.}~\bibnamefont {Nisisako}},
  \bibinfo {author} {\bibfnamefont {T.}~\bibnamefont {Torii}}, \ and\ \bibinfo
  {author} {\bibfnamefont {T.}~\bibnamefont {Higuchi}},\ }\bibfield  {title}
  {\enquote {\bibinfo {title} {Controlled production of monodisperse double
  emulsions by two-step droplet breakup in microfluidic devices},}\ }\href@noop
  {} {\bibfield  {journal} {\bibinfo  {journal} {Langmuir}\ }\textbf {\bibinfo
  {volume} {20}},\ \bibinfo {pages} {9905--9908} (\bibinfo {year}
  {2004})}\BibitemShut {NoStop}%
\bibitem [{\citenamefont {Deng}\ \emph {et~al.}(2011)\citenamefont {Deng},
  \citenamefont {Meng}, \citenamefont {Xie}, \citenamefont {Ju}, \citenamefont
  {Mou}, \citenamefont {Wang},\ and\ \citenamefont {Chu}}]{Deng2011LCWet}%
  \BibitemOpen
  \bibfield  {author} {\bibinfo {author} {\bibfnamefont {N.-N.}\ \bibnamefont
  {Deng}}, \bibinfo {author} {\bibfnamefont {Z.-J.}\ \bibnamefont {Meng}},
  \bibinfo {author} {\bibfnamefont {R.}~\bibnamefont {Xie}}, \bibinfo {author}
  {\bibfnamefont {X.-J.}\ \bibnamefont {Ju}}, \bibinfo {author} {\bibfnamefont
  {C.-L.}\ \bibnamefont {Mou}}, \bibinfo {author} {\bibfnamefont
  {W.}~\bibnamefont {Wang}}, \ and\ \bibinfo {author} {\bibfnamefont {L.-Y.}\
  \bibnamefont {Chu}},\ }\bibfield  {title} {\enquote {\bibinfo {title} {Simple
  and cheap microfluidic devices for the preparation of monodisperse
  emulsions},}\ }\href@noop {} {\bibfield  {journal} {\bibinfo  {journal} {Lab
  Chip}\ }\textbf {\bibinfo {volume} {11}},\ \bibinfo {pages} {3963--3969}
  (\bibinfo {year} {2011})}\BibitemShut {NoStop}%
\bibitem [{\citenamefont {Wan}\ \emph {et~al.}(2008)\citenamefont {Wan},
  \citenamefont {Bick}, \citenamefont {Sullivan},\ and\ \citenamefont
  {Stone}}]{Wan2008}%
  \BibitemOpen
  \bibfield  {author} {\bibinfo {author} {\bibfnamefont {J.}~\bibnamefont
  {Wan}}, \bibinfo {author} {\bibfnamefont {A.}~\bibnamefont {Bick}}, \bibinfo
  {author} {\bibfnamefont {M.}~\bibnamefont {Sullivan}}, \ and\ \bibinfo
  {author} {\bibfnamefont {H.~A.}\ \bibnamefont {Stone}},\ }\bibfield  {title}
  {\enquote {\bibinfo {title} {Controllable microfluidic production of
  microbubbles in water-in-oil emulsions and the formation of porous
  microparticles},}\ }\href@noop {} {\bibfield  {journal} {\bibinfo  {journal}
  {Adv. Mater.}\ }\textbf {\bibinfo {volume} {20}},\ \bibinfo {pages}
  {3314--3318} (\bibinfo {year} {2008})}\BibitemShut {NoStop}%
\bibitem [{\citenamefont {Che}\ \emph {et~al.}(2012)\citenamefont {Che},
  \citenamefont {Wong}, \citenamefont {Nguyen},\ and\ \citenamefont
  {Chai}}]{Che2012CPD}%
  \BibitemOpen
  \bibfield  {author} {\bibinfo {author} {\bibfnamefont {Z.}~\bibnamefont
  {Che}}, \bibinfo {author} {\bibfnamefont {T.~N.}\ \bibnamefont {Wong}},
  \bibinfo {author} {\bibfnamefont {N.-T.}\ \bibnamefont {Nguyen}}, \ and\
  \bibinfo {author} {\bibfnamefont {J.~C.}\ \bibnamefont {Chai}},\ }\bibfield
  {title} {\enquote {\bibinfo {title} {Formation and breakup of compound
  pendant drops at the tip of a capillary and its effect on upstream velocity
  fluctuations},}\ }\href@noop {} {\bibfield  {journal} {\bibinfo  {journal}
  {Int. J. Heat Mass Transfer}\ }\textbf {\bibinfo {volume} {55}},\ \bibinfo
  {pages} {1022--1029} (\bibinfo {year} {2012})}\BibitemShut {NoStop}%
\bibitem [{\citenamefont {Che}, \citenamefont {Wong},\ and\ \citenamefont
  {Nguyen}(2017)}]{Che2017compound}%
  \BibitemOpen
  \bibfield  {author} {\bibinfo {author} {\bibfnamefont {Z.}~\bibnamefont
  {Che}}, \bibinfo {author} {\bibfnamefont {T.~N.}\ \bibnamefont {Wong}}, \
  and\ \bibinfo {author} {\bibfnamefont {N.-T.}\ \bibnamefont {Nguyen}},\
  }\bibfield  {title} {\enquote {\bibinfo {title} {A simple method for the
  formation of water-in-oil-in-water ({W/O/W}) double emulsions},}\ }\href
  {\doibase 10.1007/s10404-016-1845-8} {\bibfield  {journal} {\bibinfo
  {journal} {Microfluid. Nanofluid.}\ }\textbf {\bibinfo {volume} {21}},\
  \bibinfo {pages} {8} (\bibinfo {year} {2017})}\BibitemShut {NoStop}%
\bibitem [{\citenamefont {Lorenceau}\ \emph {et~al.}(2005)\citenamefont
  {Lorenceau}, \citenamefont {Utada}, \citenamefont {Link}, \citenamefont
  {Cristobal}, \citenamefont {Joanicot},\ and\ \citenamefont
  {Weitz}}]{Lorenceau2005}%
  \BibitemOpen
  \bibfield  {author} {\bibinfo {author} {\bibfnamefont {E.}~\bibnamefont
  {Lorenceau}}, \bibinfo {author} {\bibfnamefont {A.~S.}\ \bibnamefont
  {Utada}}, \bibinfo {author} {\bibfnamefont {D.~R.}\ \bibnamefont {Link}},
  \bibinfo {author} {\bibfnamefont {G.}~\bibnamefont {Cristobal}}, \bibinfo
  {author} {\bibfnamefont {M.}~\bibnamefont {Joanicot}}, \ and\ \bibinfo
  {author} {\bibfnamefont {D.~A.}\ \bibnamefont {Weitz}},\ }\bibfield  {title}
  {\enquote {\bibinfo {title} {Generation of polymerosomes from
  double-emulsions},}\ }\href@noop {} {\bibfield  {journal} {\bibinfo
  {journal} {Langmuir}\ }\textbf {\bibinfo {volume} {21}},\ \bibinfo {pages}
  {9183--9186} (\bibinfo {year} {2005})}\BibitemShut {NoStop}%
\bibitem [{\citenamefont {Utada}\ \emph {et~al.}(2005)\citenamefont {Utada},
  \citenamefont {Lorenceau}, \citenamefont {Link}, \citenamefont {Kaplan},
  \citenamefont {Stone},\ and\ \citenamefont {Weitz}}]{Utada2005}%
  \BibitemOpen
  \bibfield  {author} {\bibinfo {author} {\bibfnamefont {A.~S.}\ \bibnamefont
  {Utada}}, \bibinfo {author} {\bibfnamefont {E.}~\bibnamefont {Lorenceau}},
  \bibinfo {author} {\bibfnamefont {D.~R.}\ \bibnamefont {Link}}, \bibinfo
  {author} {\bibfnamefont {P.~D.}\ \bibnamefont {Kaplan}}, \bibinfo {author}
  {\bibfnamefont {H.~A.}\ \bibnamefont {Stone}}, \ and\ \bibinfo {author}
  {\bibfnamefont {D.~A.}\ \bibnamefont {Weitz}},\ }\bibfield  {title} {\enquote
  {\bibinfo {title} {Monodisperse double emulsions generated from a
  microcapillary device},}\ }\href@noop {} {\bibfield  {journal} {\bibinfo
  {journal} {Science}\ }\textbf {\bibinfo {volume} {308}},\ \bibinfo {pages}
  {537--541} (\bibinfo {year} {2005})}\BibitemShut {NoStop}%
\bibitem [{\citenamefont {Azarmanesh}, \citenamefont {Farhadi},\ and\
  \citenamefont {Azizian}(2016)}]{Azarmanesh2016DoubleEmulsionGerris}%
  \BibitemOpen
  \bibfield  {author} {\bibinfo {author} {\bibfnamefont {M.}~\bibnamefont
  {Azarmanesh}}, \bibinfo {author} {\bibfnamefont {M.}~\bibnamefont {Farhadi}},
  \ and\ \bibinfo {author} {\bibfnamefont {P.}~\bibnamefont {Azizian}},\
  }\bibfield  {title} {\enquote {\bibinfo {title} {Double emulsion formation
  through hierarchical flow-focusing microchannel},}\ }\href@noop {} {\bibfield
   {journal} {\bibinfo  {journal} {Phys. Fluids}\ }\textbf {\bibinfo {volume}
  {28}},\ \bibinfo {pages} {032005} (\bibinfo {year} {2016})}\BibitemShut
  {NoStop}%
\bibitem [{\citenamefont {Zhou}, \citenamefont {Yue},\ and\ \citenamefont
  {Feng}(2006)}]{Zhou2006}%
  \BibitemOpen
  \bibfield  {author} {\bibinfo {author} {\bibfnamefont {C.}~\bibnamefont
  {Zhou}}, \bibinfo {author} {\bibfnamefont {P.}~\bibnamefont {Yue}}, \ and\
  \bibinfo {author} {\bibfnamefont {J.~J.}\ \bibnamefont {Feng}},\ }\bibfield
  {title} {\enquote {\bibinfo {title} {Formation of simple and compound drops
  in microfluidic devices},}\ }\href@noop {} {\bibfield  {journal} {\bibinfo
  {journal} {Phys. Fluids}\ }\textbf {\bibinfo {volume} {18}},\ \bibinfo
  {pages} {092105} (\bibinfo {year} {2006})}\BibitemShut {NoStop}%
\bibitem [{\citenamefont {Nabavi}\ \emph {et~al.}(2015)\citenamefont {Nabavi},
  \citenamefont {Vladisavljevi{\'c}}, \citenamefont {Gu},\ and\ \citenamefont
  {Ekanem}}]{nabavi2015doubleCFD}%
  \BibitemOpen
  \bibfield  {author} {\bibinfo {author} {\bibfnamefont {S.~A.}\ \bibnamefont
  {Nabavi}}, \bibinfo {author} {\bibfnamefont {G.~T.}\ \bibnamefont
  {Vladisavljevi{\'c}}}, \bibinfo {author} {\bibfnamefont {S.}~\bibnamefont
  {Gu}}, \ and\ \bibinfo {author} {\bibfnamefont {E.~E.}\ \bibnamefont
  {Ekanem}},\ }\bibfield  {title} {\enquote {\bibinfo {title} {Double emulsion
  production in glass capillary microfluidic device: Parametric investigation
  of droplet generation behaviour},}\ }\href@noop {} {\bibfield  {journal}
  {\bibinfo  {journal} {Chem. Eng. Sci.}\ }\textbf {\bibinfo {volume} {130}},\
  \bibinfo {pages} {183--196} (\bibinfo {year} {2015})}\BibitemShut {NoStop}%
\bibitem [{\citenamefont {Stone}\ and\ \citenamefont
  {Leal}(1990)}]{Stone1990CompoundLinearFlow}%
  \BibitemOpen
  \bibfield  {author} {\bibinfo {author} {\bibfnamefont {H.~A.}\ \bibnamefont
  {Stone}}\ and\ \bibinfo {author} {\bibfnamefont {L.~G.}\ \bibnamefont
  {Leal}},\ }\bibfield  {title} {\enquote {\bibinfo {title} {Breakup of
  concentric double emulsion droplets in linear flows},}\ }\href@noop {}
  {\bibfield  {journal} {\bibinfo  {journal} {J. Fluid. Mech.}\ }\textbf
  {\bibinfo {volume} {211}},\ \bibinfo {pages} {123--156} (\bibinfo {year}
  {1990})}\BibitemShut {NoStop}%
\bibitem [{\citenamefont {Qu}\ and\ \citenamefont
  {Wang}(2012)}]{Qu2012CompoundExtensionalFlow}%
  \BibitemOpen
  \bibfield  {author} {\bibinfo {author} {\bibfnamefont {X.}~\bibnamefont
  {Qu}}\ and\ \bibinfo {author} {\bibfnamefont {Y.}~\bibnamefont {Wang}},\
  }\bibfield  {title} {\enquote {\bibinfo {title} {Dynamics of concentric and
  eccentric compound droplets suspended in extensional flows},}\ }\href@noop {}
  {\bibfield  {journal} {\bibinfo  {journal} {Phys. Fluids}\ }\textbf {\bibinfo
  {volume} {24}},\ \bibinfo {pages} {123302} (\bibinfo {year}
  {2012})}\BibitemShut {NoStop}%
\bibitem [{\citenamefont {Hua}, \citenamefont {Shin},\ and\ \citenamefont
  {Kim}(2014)}]{Hua2014CompoundShearFlow}%
  \BibitemOpen
  \bibfield  {author} {\bibinfo {author} {\bibfnamefont {H.}~\bibnamefont
  {Hua}}, \bibinfo {author} {\bibfnamefont {J.}~\bibnamefont {Shin}}, \ and\
  \bibinfo {author} {\bibfnamefont {J.}~\bibnamefont {Kim}},\ }\bibfield
  {title} {\enquote {\bibinfo {title} {Dynamics of a compound droplet in shear
  flow},}\ }\href@noop {} {\bibfield  {journal} {\bibinfo  {journal} {Int. J.
  Heat Fluid. Fl.}\ }\textbf {\bibinfo {volume} {50}},\ \bibinfo {pages}
  {63--71} (\bibinfo {year} {2014})}\BibitemShut {NoStop}%
\bibitem [{\citenamefont {Smith}, \citenamefont {Ottino},\ and\ \citenamefont
  {Olvera de~la Cruz}(2004)}]{Smith2004CompoundShearFlowPRL}%
  \BibitemOpen
  \bibfield  {author} {\bibinfo {author} {\bibfnamefont {K.~A.}\ \bibnamefont
  {Smith}}, \bibinfo {author} {\bibfnamefont {J.~M.}\ \bibnamefont {Ottino}}, \
  and\ \bibinfo {author} {\bibfnamefont {M.}~\bibnamefont {Olvera de~la
  Cruz}},\ }\bibfield  {title} {\enquote {\bibinfo {title} {Encapsulated drop
  breakup in shear flow},}\ }\href@noop {} {\bibfield  {journal} {\bibinfo
  {journal} {Phys. Rev. Lett.}\ }\textbf {\bibinfo {volume} {93}},\ \bibinfo
  {pages} {204501} (\bibinfo {year} {2004})}\BibitemShut {NoStop}%
\bibitem [{\citenamefont {Zhou}, \citenamefont {Yue},\ and\ \citenamefont
  {Feng}(2008)}]{Zhou2008CompoundContraction}%
  \BibitemOpen
  \bibfield  {author} {\bibinfo {author} {\bibfnamefont {C.}~\bibnamefont
  {Zhou}}, \bibinfo {author} {\bibfnamefont {P.}~\bibnamefont {Yue}}, \ and\
  \bibinfo {author} {\bibfnamefont {J.~J.}\ \bibnamefont {Feng}},\ }\bibfield
  {title} {\enquote {\bibinfo {title} {Deformation of a compound drop through a
  contraction in a pressure-driven pipe flow},}\ }\href@noop {} {\bibfield
  {journal} {\bibinfo  {journal} {Int. J. Multiphase Flow}\ }\textbf {\bibinfo
  {volume} {34}},\ \bibinfo {pages} {102 -- 109} (\bibinfo {year}
  {2008})}\BibitemShut {NoStop}%
\bibitem [{\citenamefont {Tao}\ \emph {et~al.}(2013)\citenamefont {Tao},
  \citenamefont {Song}, \citenamefont {Liu},\ and\ \citenamefont
  {Wang}}]{Tao2013CompoundContraction}%
  \BibitemOpen
  \bibfield  {author} {\bibinfo {author} {\bibfnamefont {J.}~\bibnamefont
  {Tao}}, \bibinfo {author} {\bibfnamefont {X.}~\bibnamefont {Song}}, \bibinfo
  {author} {\bibfnamefont {J.}~\bibnamefont {Liu}}, \ and\ \bibinfo {author}
  {\bibfnamefont {J.}~\bibnamefont {Wang}},\ }\bibfield  {title} {\enquote
  {\bibinfo {title} {Microfluidic rheology of the multiple-emulsion globule
  transiting in a contraction tube through a boundary element method},}\
  }\href@noop {} {\bibfield  {journal} {\bibinfo  {journal} {Chem. Eng. Sci.}\
  }\textbf {\bibinfo {volume} {97}},\ \bibinfo {pages} {328 -- 336} (\bibinfo
  {year} {2013})}\BibitemShut {NoStop}%
\bibitem [{\citenamefont {Song}, \citenamefont {Xu},\ and\ \citenamefont
  {Yang}(2010)}]{Song2010PoFCompound}%
  \BibitemOpen
  \bibfield  {author} {\bibinfo {author} {\bibfnamefont {Y.}~\bibnamefont
  {Song}}, \bibinfo {author} {\bibfnamefont {J.}~\bibnamefont {Xu}}, \ and\
  \bibinfo {author} {\bibfnamefont {Y.}~\bibnamefont {Yang}},\ }\bibfield
  {title} {\enquote {\bibinfo {title} {Stokes flow past a compound drop in a
  circular tube},}\ }\href@noop {} {\bibfield  {journal} {\bibinfo  {journal}
  {Phys. Fluids}\ }\textbf {\bibinfo {volume} {22}},\ \bibinfo {pages} {072003}
  (\bibinfo {year} {2010})}\BibitemShut {NoStop}%
\bibitem [{\citenamefont {Chu}\ \emph {et~al.}(2007)\citenamefont {Chu},
  \citenamefont {Utada}, \citenamefont {Shah}, \citenamefont {Kim},\ and\
  \citenamefont {Weitz}}]{Chu2007HighEmulsion}%
  \BibitemOpen
  \bibfield  {author} {\bibinfo {author} {\bibfnamefont {L.-Y.}\ \bibnamefont
  {Chu}}, \bibinfo {author} {\bibfnamefont {A.~S.}\ \bibnamefont {Utada}},
  \bibinfo {author} {\bibfnamefont {R.~K.}\ \bibnamefont {Shah}}, \bibinfo
  {author} {\bibfnamefont {J.-W.}\ \bibnamefont {Kim}}, \ and\ \bibinfo
  {author} {\bibfnamefont {D.~A.}\ \bibnamefont {Weitz}},\ }\bibfield  {title}
  {\enquote {\bibinfo {title} {Controllable monodisperse multiple emulsions},}\
  }\href@noop {} {\bibfield  {journal} {\bibinfo  {journal} {Angew. Chem. Int.
  Ed.}\ }\textbf {\bibinfo {volume} {46}},\ \bibinfo {pages} {8970--8974}
  (\bibinfo {year} {2007})}\BibitemShut {NoStop}%
\bibitem [{\citenamefont {Wang}\ \emph {et~al.}(2011)\citenamefont {Wang},
  \citenamefont {Xie}, \citenamefont {Ju}, \citenamefont {Luo}, \citenamefont
  {Liu}, \citenamefont {Weitz},\ and\ \citenamefont
  {Chu}}]{Wang2011HighEmulsion}%
  \BibitemOpen
  \bibfield  {author} {\bibinfo {author} {\bibfnamefont {W.}~\bibnamefont
  {Wang}}, \bibinfo {author} {\bibfnamefont {R.}~\bibnamefont {Xie}}, \bibinfo
  {author} {\bibfnamefont {X.-J.}\ \bibnamefont {Ju}}, \bibinfo {author}
  {\bibfnamefont {T.}~\bibnamefont {Luo}}, \bibinfo {author} {\bibfnamefont
  {L.}~\bibnamefont {Liu}}, \bibinfo {author} {\bibfnamefont {D.~A.}\
  \bibnamefont {Weitz}}, \ and\ \bibinfo {author} {\bibfnamefont {L.-Y.}\
  \bibnamefont {Chu}},\ }\bibfield  {title} {\enquote {\bibinfo {title}
  {Controllable microfluidic production of multicomponent multiple
  emulsions},}\ }\href@noop {} {\bibfield  {journal} {\bibinfo  {journal} {Lab
  Chip}\ }\textbf {\bibinfo {volume} {11}},\ \bibinfo {pages} {1587--1592}
  (\bibinfo {year} {2011})}\BibitemShut {NoStop}%
\bibitem [{\citenamefont {Abate}\ and\ \citenamefont
  {Weitz}(2009)}]{Abate2009HighEmulsion}%
  \BibitemOpen
  \bibfield  {author} {\bibinfo {author} {\bibfnamefont {A.~R.}\ \bibnamefont
  {Abate}}\ and\ \bibinfo {author} {\bibfnamefont {D.~A.}\ \bibnamefont
  {Weitz}},\ }\bibfield  {title} {\enquote {\bibinfo {title} {High-order
  multiple emulsions formed in poly (dimethylsiloxane) microfluidics},}\
  }\href@noop {} {\bibfield  {journal} {\bibinfo  {journal} {Small}\ }\textbf
  {\bibinfo {volume} {5}},\ \bibinfo {pages} {2030--2032} (\bibinfo {year}
  {2009})}\BibitemShut {NoStop}%
\bibitem [{\citenamefont {Kim}\ and\ \citenamefont
  {Weitz}(2011)}]{Kim2011HighEmulsion}%
  \BibitemOpen
  \bibfield  {author} {\bibinfo {author} {\bibfnamefont {S.-H.}\ \bibnamefont
  {Kim}}\ and\ \bibinfo {author} {\bibfnamefont {D.~A.}\ \bibnamefont
  {Weitz}},\ }\bibfield  {title} {\enquote {\bibinfo {title} {One-step
  emulsification of multiple concentric shells with capillary microfluidic
  devices},}\ }\href@noop {} {\bibfield  {journal} {\bibinfo  {journal} {Angew.
  Chem. Int. Ed.}\ }\textbf {\bibinfo {volume} {50}},\ \bibinfo {pages}
  {8731--8734} (\bibinfo {year} {2011})}\BibitemShut {NoStop}%
\bibitem [{\citenamefont {Ralf}\ \emph {et~al.}(2012)\citenamefont {Ralf},
  \citenamefont {Martin}, \citenamefont {Thomas},\ and\ \citenamefont
  {Stephan}}]{Ralf2012}%
  \BibitemOpen
  \bibfield  {author} {\bibinfo {author} {\bibfnamefont {S.}~\bibnamefont
  {Ralf}}, \bibinfo {author} {\bibfnamefont {B.}~\bibnamefont {Martin}},
  \bibinfo {author} {\bibfnamefont {P.}~\bibnamefont {Thomas}}, \ and\ \bibinfo
  {author} {\bibfnamefont {H.}~\bibnamefont {Stephan}},\ }\bibfield  {title}
  {\enquote {\bibinfo {title} {Droplet based microfluidics},}\ }\href@noop {}
  {\bibfield  {journal} {\bibinfo  {journal} {Reports on Progress in Physics}\
  }\textbf {\bibinfo {volume} {75}},\ \bibinfo {pages} {016601} (\bibinfo
  {year} {2012})}\BibitemShut {NoStop}%
\bibitem [{\citenamefont {Patankar}(1980)}]{Patankar1980}%
  \BibitemOpen
  \bibfield  {author} {\bibinfo {author} {\bibfnamefont {S.~V.}\ \bibnamefont
  {Patankar}},\ }\href@noop {} {\emph {\bibinfo {title} {Numerical Heat
  Transfer and Fluid Flow}}},\ Series in Computational Methods in Mechanics and
  Thermal Sciences\ (\bibinfo  {publisher} {Hemisphere Pub. Corp.},\ \bibinfo
  {address} {Washington},\ \bibinfo {year} {1980})\ pp.\ \bibinfo {pages}
  {xiii, 197 p.}\BibitemShut {Stop}%
\bibitem [{\citenamefont {Che}\ \emph {et~al.}(2011)\citenamefont {Che},
  \citenamefont {Wong}, \citenamefont {Nguyen}, \citenamefont {Yap},\ and\
  \citenamefont {Chai}}]{Che2011Pendent}%
  \BibitemOpen
  \bibfield  {author} {\bibinfo {author} {\bibfnamefont {Z.}~\bibnamefont
  {Che}}, \bibinfo {author} {\bibfnamefont {T.~N.}\ \bibnamefont {Wong}},
  \bibinfo {author} {\bibfnamefont {N.-T.}\ \bibnamefont {Nguyen}}, \bibinfo
  {author} {\bibfnamefont {Y.~F.}\ \bibnamefont {Yap}}, \ and\ \bibinfo
  {author} {\bibfnamefont {J.}~\bibnamefont {Chai}},\ }\bibfield  {title}
  {\enquote {\bibinfo {title} {Numerical investigation of upstream pressure
  fluctuation during growth and breakup of pendant drops},}\ }\href@noop {}
  {\bibfield  {journal} {\bibinfo  {journal} {Chem. Eng. Sci.}\ }\textbf
  {\bibinfo {volume} {66}},\ \bibinfo {pages} {5293--5300} (\bibinfo {year}
  {2011})}\BibitemShut {NoStop}%
\bibitem [{\citenamefont {Kang}, \citenamefont {Fedkiw},\ and\ \citenamefont
  {Liu}(2000)}]{Kang2000}%
  \BibitemOpen
  \bibfield  {author} {\bibinfo {author} {\bibfnamefont {M.}~\bibnamefont
  {Kang}}, \bibinfo {author} {\bibfnamefont {R.~P.}\ \bibnamefont {Fedkiw}}, \
  and\ \bibinfo {author} {\bibfnamefont {X.~D.}\ \bibnamefont {Liu}},\
  }\bibfield  {title} {\enquote {\bibinfo {title} {A boundary condition
  capturing method for multiphase incompressible flow},}\ }\href@noop {}
  {\bibfield  {journal} {\bibinfo  {journal} {J. Sci. Comput.}\ }\textbf
  {\bibinfo {volume} {15}},\ \bibinfo {pages} {323--360} (\bibinfo {year}
  {2000})}\BibitemShut {NoStop}%
\bibitem [{\citenamefont {Liu}, \citenamefont {Fedkiw},\ and\ \citenamefont
  {Kang}(2000)}]{Liu2000}%
  \BibitemOpen
  \bibfield  {author} {\bibinfo {author} {\bibfnamefont {X.~D.}\ \bibnamefont
  {Liu}}, \bibinfo {author} {\bibfnamefont {R.~P.}\ \bibnamefont {Fedkiw}}, \
  and\ \bibinfo {author} {\bibfnamefont {M.}~\bibnamefont {Kang}},\ }\bibfield
  {title} {\enquote {\bibinfo {title} {A boundary condition capturing method
  for {P}oisson's equation on irregular domains},}\ }\href@noop {} {\bibfield
  {journal} {\bibinfo  {journal} {J. Comput. Phys.}\ }\textbf {\bibinfo
  {volume} {160}},\ \bibinfo {pages} {151--178} (\bibinfo {year}
  {2000})}\BibitemShut {NoStop}%
\bibitem [{\citenamefont {Osher}\ and\ \citenamefont
  {Fedkiw}(2003)}]{Osher2003}%
  \BibitemOpen
  \bibfield  {author} {\bibinfo {author} {\bibfnamefont {S.}~\bibnamefont
  {Osher}}\ and\ \bibinfo {author} {\bibfnamefont {R.~P.}\ \bibnamefont
  {Fedkiw}},\ }\href@noop {} {\emph {\bibinfo {title} {Level Set Methods and
  Dynamic Implicit Surfaces}}},\ Applied Mathematical Sciences\ (\bibinfo
  {publisher} {Springer},\ \bibinfo {address} {New York},\ \bibinfo {year}
  {2003})\ pp.\ \bibinfo {pages} {xii, 273 p.}\BibitemShut {Stop}%
\bibitem [{\citenamefont {Shu}\ and\ \citenamefont {Osher}(1988)}]{Shu1988RK3}%
  \BibitemOpen
  \bibfield  {author} {\bibinfo {author} {\bibfnamefont {C.-W.}\ \bibnamefont
  {Shu}}\ and\ \bibinfo {author} {\bibfnamefont {S.}~\bibnamefont {Osher}},\
  }\bibfield  {title} {\enquote {\bibinfo {title} {Efficient implementation of
  essentially non-oscillatory shock-capturing schemes},}\ }\href@noop {}
  {\bibfield  {journal} {\bibinfo  {journal} {J. Comput. Phys.}\ }\textbf
  {\bibinfo {volume} {77}},\ \bibinfo {pages} {439--471} (\bibinfo {year}
  {1988})}\BibitemShut {NoStop}%
\bibitem [{\citenamefont {Tryggvason}, \citenamefont {Scardovelli},\ and\
  \citenamefont {Zaleski}(2011)}]{Tryggvason2011book}%
  \BibitemOpen
  \bibfield  {author} {\bibinfo {author} {\bibfnamefont {G.}~\bibnamefont
  {Tryggvason}}, \bibinfo {author} {\bibfnamefont {R.}~\bibnamefont
  {Scardovelli}}, \ and\ \bibinfo {author} {\bibfnamefont {S.}~\bibnamefont
  {Zaleski}},\ }\href@noop {} {\emph {\bibinfo {title} {Direct numerical
  simulations of gas–liquid multiphase flows}}}\ (\bibinfo  {publisher}
  {Cambridge University Press},\ \bibinfo {year} {2011})\BibitemShut {NoStop}%
\bibitem [{\citenamefont {Dendukuri}\ and\ \citenamefont
  {Doyle}(2009)}]{Dendukuri2009ParticleReview}%
  \BibitemOpen
  \bibfield  {author} {\bibinfo {author} {\bibfnamefont {D.}~\bibnamefont
  {Dendukuri}}\ and\ \bibinfo {author} {\bibfnamefont {P.~S.}\ \bibnamefont
  {Doyle}},\ }\bibfield  {title} {\enquote {\bibinfo {title} {The synthesis and
  assembly of polymeric microparticles using microfluidics},}\ }\href@noop {}
  {\bibfield  {journal} {\bibinfo  {journal} {Adv. Mater.}\ }\textbf {\bibinfo
  {volume} {21}},\ \bibinfo {pages} {4071--4086} (\bibinfo {year}
  {2009})}\BibitemShut {NoStop}%
\bibitem [{\citenamefont {Park}\ \emph {et~al.}(2010)\citenamefont {Park},
  \citenamefont {Saffari}, \citenamefont {Kumar}, \citenamefont {G\"{u}nther},\
  and\ \citenamefont {Kumacheva}}]{Kumacheva2010PoymerizationReview}%
  \BibitemOpen
  \bibfield  {author} {\bibinfo {author} {\bibfnamefont {J.~I.}\ \bibnamefont
  {Park}}, \bibinfo {author} {\bibfnamefont {A.}~\bibnamefont {Saffari}},
  \bibinfo {author} {\bibfnamefont {S.}~\bibnamefont {Kumar}}, \bibinfo
  {author} {\bibfnamefont {A.}~\bibnamefont {G\"{u}nther}}, \ and\ \bibinfo
  {author} {\bibfnamefont {E.}~\bibnamefont {Kumacheva}},\ }\bibfield  {title}
  {\enquote {\bibinfo {title} {Microfluidic synthesis of polymer and inorganic
  particulate materials},}\ }\href@noop {} {\bibfield  {journal} {\bibinfo
  {journal} {Ann. Rev. Mater. Res.}\ }\textbf {\bibinfo {volume} {40}},\
  \bibinfo {pages} {415--443} (\bibinfo {year} {2010})}\BibitemShut {NoStop}%
\bibitem [{\citenamefont {Lee}\ \emph {et~al.}(2016)\citenamefont {Lee},
  \citenamefont {Choi}, \citenamefont {Abbaspourrad}, \citenamefont {Wesner},
  \citenamefont {Caggioni}, \citenamefont {Zhu}, \citenamefont {Nawar},\ and\
  \citenamefont {Weitz}}]{Lee2016TripleWeitz}%
  \BibitemOpen
  \bibfield  {author} {\bibinfo {author} {\bibfnamefont {H.}~\bibnamefont
  {Lee}}, \bibinfo {author} {\bibfnamefont {C.-H.}\ \bibnamefont {Choi}},
  \bibinfo {author} {\bibfnamefont {A.}~\bibnamefont {Abbaspourrad}}, \bibinfo
  {author} {\bibfnamefont {C.}~\bibnamefont {Wesner}}, \bibinfo {author}
  {\bibfnamefont {M.}~\bibnamefont {Caggioni}}, \bibinfo {author}
  {\bibfnamefont {T.}~\bibnamefont {Zhu}}, \bibinfo {author} {\bibfnamefont
  {S.}~\bibnamefont {Nawar}}, \ and\ \bibinfo {author} {\bibfnamefont {D.~A.}\
  \bibnamefont {Weitz}},\ }\bibfield  {title} {\enquote {\bibinfo {title}
  {Fluorocarbon oil reinforced triple emulsion drops},}\ }\href {\doibase
  10.1002/adma.201602804} {\bibfield  {journal} {\bibinfo  {journal} {Adv.
  Mater.}\ ,\ \bibinfo {pages} {8425--8430}} (\bibinfo {year}
  {2016})}\BibitemShut {NoStop}%
\bibitem [{\citenamefont {Bandara}, \citenamefont {Nguyen},\ and\ \citenamefont
  {Rosengarten}(2015)}]{Bandara2015DropletHeatTransferReview}%
  \BibitemOpen
  \bibfield  {author} {\bibinfo {author} {\bibfnamefont {T.}~\bibnamefont
  {Bandara}}, \bibinfo {author} {\bibfnamefont {N.-T.}\ \bibnamefont {Nguyen}},
  \ and\ \bibinfo {author} {\bibfnamefont {G.}~\bibnamefont {Rosengarten}},\
  }\bibfield  {title} {\enquote {\bibinfo {title} {Slug flow heat transfer
  without phase change in microchannels: A review},}\ }\href@noop {} {\bibfield
   {journal} {\bibinfo  {journal} {Chem. Eng. Sci.}\ }\textbf {\bibinfo
  {volume} {126}},\ \bibinfo {pages} {283--295} (\bibinfo {year}
  {2015})}\BibitemShut {NoStop}%
\bibitem [{\citenamefont {Che}, \citenamefont {Wong},\ and\ \citenamefont
  {Nguyen}(2012)}]{Che2012HeatTransfer2DTA}%
  \BibitemOpen
  \bibfield  {author} {\bibinfo {author} {\bibfnamefont {Z.}~\bibnamefont
  {Che}}, \bibinfo {author} {\bibfnamefont {T.~N.}\ \bibnamefont {Wong}}, \
  and\ \bibinfo {author} {\bibfnamefont {N.-T.}\ \bibnamefont {Nguyen}},\
  }\bibfield  {title} {\enquote {\bibinfo {title} {Heat transfer enhancement by
  recirculating flow within liquid plugs in microchannels},}\ }\href@noop {}
  {\bibfield  {journal} {\bibinfo  {journal} {Int. J. Heat Mass. Tran.}\
  }\textbf {\bibinfo {volume} {55}},\ \bibinfo {pages} {1947--1956} (\bibinfo
  {year} {2012})}\BibitemShut {NoStop}%
\bibitem [{\citenamefont {Che}\ \emph {et~al.}(2015)\citenamefont {Che},
  \citenamefont {Wong}, \citenamefont {Nguyen},\ and\ \citenamefont
  {Yang}}]{Che2015HeatTransfer3Ddroplet}%
  \BibitemOpen
  \bibfield  {author} {\bibinfo {author} {\bibfnamefont {Z.}~\bibnamefont
  {Che}}, \bibinfo {author} {\bibfnamefont {T.~N.}\ \bibnamefont {Wong}},
  \bibinfo {author} {\bibfnamefont {N.-T.}\ \bibnamefont {Nguyen}}, \ and\
  \bibinfo {author} {\bibfnamefont {C.}~\bibnamefont {Yang}},\ }\bibfield
  {title} {\enquote {\bibinfo {title} {Three dimensional features of convective
  heat transfer in droplet-based microchannel heat sinks},}\ }\href@noop {}
  {\bibfield  {journal} {\bibinfo  {journal} {Int. J. Heat Mass. Tran.}\
  }\textbf {\bibinfo {volume} {86}},\ \bibinfo {pages} {455--464} (\bibinfo
  {year} {2015})}\BibitemShut {NoStop}%
\bibitem [{\citenamefont {Handique}\ and\ \citenamefont
  {Burns}(2001)}]{Handique2001}%
  \BibitemOpen
  \bibfield  {author} {\bibinfo {author} {\bibfnamefont {K.}~\bibnamefont
  {Handique}}\ and\ \bibinfo {author} {\bibfnamefont {M.~A.}\ \bibnamefont
  {Burns}},\ }\bibfield  {title} {\enquote {\bibinfo {title} {Mathematical
  modeling of drop mixing in a slit-type microchannel},}\ }\href@noop {}
  {\bibfield  {journal} {\bibinfo  {journal} {J. Micromech. Microeng}\ }\textbf
  {\bibinfo {volume} {11}},\ \bibinfo {pages} {548--554} (\bibinfo {year}
  {2001})}\BibitemShut {NoStop}%
\bibitem [{\citenamefont {Tice}\ \emph {et~al.}(2003)\citenamefont {Tice},
  \citenamefont {Song}, \citenamefont {Lyon},\ and\ \citenamefont
  {Ismagilov}}]{Tice2003DropletMixing}%
  \BibitemOpen
  \bibfield  {author} {\bibinfo {author} {\bibfnamefont {J.~D.}\ \bibnamefont
  {Tice}}, \bibinfo {author} {\bibfnamefont {H.}~\bibnamefont {Song}}, \bibinfo
  {author} {\bibfnamefont {A.~D.}\ \bibnamefont {Lyon}}, \ and\ \bibinfo
  {author} {\bibfnamefont {R.~F.}\ \bibnamefont {Ismagilov}},\ }\bibfield
  {title} {\enquote {\bibinfo {title} {Formation of droplets and mixing in
  multiphase microfluidics at low values of the reynolds and the capillary
  numbers},}\ }\href@noop {} {\bibfield  {journal} {\bibinfo  {journal}
  {Langmuir}\ }\textbf {\bibinfo {volume} {19}},\ \bibinfo {pages} {9127--9133}
  (\bibinfo {year} {2003})}\BibitemShut {NoStop}%
\bibitem [{\citenamefont {Che}, \citenamefont {Nguyen},\ and\ \citenamefont
  {Wong}(2011)}]{Che2011ChaoticMixingPRE}%
  \BibitemOpen
  \bibfield  {author} {\bibinfo {author} {\bibfnamefont {Z.}~\bibnamefont
  {Che}}, \bibinfo {author} {\bibfnamefont {N.-T.}\ \bibnamefont {Nguyen}}, \
  and\ \bibinfo {author} {\bibfnamefont {T.~N.}\ \bibnamefont {Wong}},\
  }\bibfield  {title} {\enquote {\bibinfo {title} {Analysis of chaotic mixing
  in plugs moving in meandering microchannels},}\ }\href@noop {} {\bibfield
  {journal} {\bibinfo  {journal} {Phys. Rev. E}\ }\textbf {\bibinfo {volume}
  {84}},\ \bibinfo {pages} {066309} (\bibinfo {year} {2011})}\BibitemShut
  {NoStop}%
\end{thebibliography}%
\end{document}